\documentclass[runningheads]{llncs}
\usepackage{graphicx}
\usepackage{comment}
\usepackage{amsmath,amssymb} 
\usepackage{color}
\usepackage{mydefs}
\usepackage{url}

\newcommand*{\eg}{\textit{e.g.}}
\newcommand*{\ie}{\textit{i.e.}}
\newcommand*{\etal}{\textit{et al.}}

\begin{document}
\pagestyle{headings}
\mainmatter
\def\ECCVSubNumber{47}  

\title{Learning to Improve Image Compression \\  without Changing the Standard Decoder} 

\titlerunning{Learned Image Compression without Changing the Standard Decoder}
%
\author{Yannick Str\"umpler \orcidID{0000-0002-9810-5112} \and
Ren Yang\orcidID{0000-0003-4124-4186} \and
Radu Timofte \orcidID{0000-0002-1478-0402}}
\authorrunning{Y. Str\"umpler et al.}
%
\institute{ETH Zurich, R\"amistrasse 101, 8092 Zurich, Switzerland \\
\email{styannic@ethz.ch, \{ren.yang, timofter\}@vision.ee.ethz.ch}}
\maketitle

\begin{abstract}
In recent years we have witnessed an increasing interest in applying Deep Neural Networks (DNNs) to improve the rate-distortion performance in image compression. However, the existing approaches either train a post-processing DNN on the decoder side, or propose learning for image compression in an end-to-end manner. This way, the trained DNNs are required in the decoder, leading to the incompatibility to the standard image decoders (\eg, JPEG) in personal computers and mobiles. Therefore, we propose learning to improve the encoding performance with the standard decoder. In this paper, We work on JPEG as an example. Specifically, a frequency-domain pre-editing method is proposed to optimize the distribution of DCT coefficients, aiming at facilitating the JPEG compression. Moreover, we propose learning the JPEG quantization table jointly with the pre-editing network. Most importantly, we do not modify the JPEG decoder and therefore our approach is applicable when viewing images with the widely used standard JPEG decoder. The experiments validate that our approach successfully improves the rate-distortion performance of JPEG in terms of various quality metrics, such as PSNR, MS-SSIM and LPIPS. Visually, this translates to better overall color retention especially when strong compression is applied. The codes are available at \url{https://github.com/YannickStruempler/LearnedJPEG}.
\end{abstract}

\section{Introduction}
The past decades have witnessed the increasing popularity of transmitting images over the Internet, while also the typical image resolution has become larger. Therefore, improving the performance of image compression is essential for the efficient transmission of images over the band-limited Internet. In recent years, there has been an increasing interest in employing Deep Neural Networks (DNNs) into image compression frameworks to improve the rate-distortion performance. Specifically, some works, \eg, \cite{dong2015compression,wang2016d3,zhang2017beyond,tai2017memnet,zhang2020residual}, apply DNNs to reduce the compression artifacts on the decoder side. Other works, \eg, \cite{Toderici2016Variable,balle2017end,balle2018variational,mentzer2018conditional,lee2019context,Hu2020Coarse}, proposed learning for image compression with end-to-end DNNs and advance the state-of-the-art performance of image compression. However, each of these approaches requires a specifically trained decoder (post-processing DNNs or the DNN-based decoder), and therefore cannot be supported by the commonly used image viewers in computers and mobiles. Such incompatibility reduces the practicability of these approaches. 

\begin{figure}[!t]
\centering
\includegraphics[width=\linewidth]{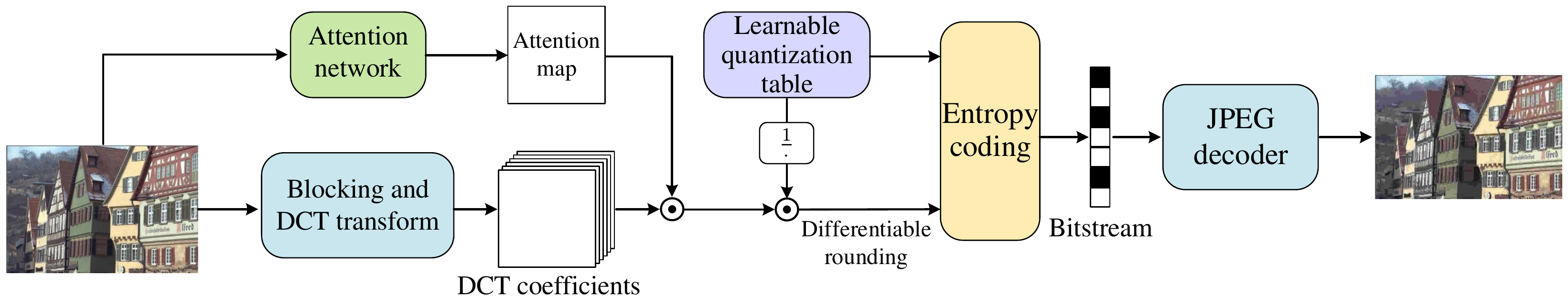}
\caption{The overview of the proposed approach. In this paper, $\odot$ indicates element-wise multiplication.}
\label{fig:1}
\end{figure}

To overcome this shortcoming, this paper proposes adopting deep learning strategies to optimize the handcrafted image encoder without modification in the decoder side. We work on the most commonly used image compression standard JPEG~\cite{jpeg} as an example. Our approach improves the rate-distortion performance of JPEG while ensuring that the bitstreams are decodable by the standard JPEG decoder. As such, it is compatible with all image viewers in personal computers and mobiles. To be specific, as shown in Figure~\ref{fig:1}, we propose pre-editing the input image in the frequency domain by a learned attention map. The attention map learns to weight the DCT coefficients to facilitate the compression of the input image. Moreover, we propose learning the quantization table in the JPEG encoder. Unlike the standard JPEG that uses hand-crafted quantization tables, we propose jointly optimizing them with the attention network for rate-distortion performance. Note that, since the DCT transform is differentiable, we build a differentiable JPEG pipeline during training, and thus the proposed attention network and the learnable quantization table can be jointly trained in an end-to-end manner. 


The contribution of this paper can be summarized as:
\begin{enumerate}
    \item We propose improving the rate-distortion performance by optimizing the JPEG encoder in the frequency domain and keep the standard JPEG decoder.
    \item We propose an attention network which learns to facilitate the JPEG compression by editing the DCT coefficients.
    \item We propose a learnable quantization table which can be jointly optimized with the attention network towards the rate-distortion performance in an end-to-end manner.
\end{enumerate}

\section{Related work}

During the past a few years, plenty of works apply DNNs to improve the performance of image compression. Among them, \cite{dong2015compression,guo2016building,wang2016d3,zhang2017beyond,tai2017memnet,li2017efficient,dai2017convolutional,yang2017decoder,yang2018enhancing,xing2020early} proposed post-processing DNNs to enhance the compressed image without bit-rate overhead, thus improving the rate-distortion performance. For example, Dong~\etal~\cite{dong2015compression} proposed a four-layer DNN to reduce the compression artifacts of JPEG. Later, Guo~\etal~\cite{guo2016building} and Wang~\etal~\cite{wang2016d3} designed the advanced post-processing network based on the prior knowledge of the JPEG algorithm. Afterwards, DnCNN~\cite{zhang2017beyond} and Memnet~\cite{tai2017memnet} were proposed for various restoration tasks, including the enhancement of JPEG images. Meanwhile, some approaches aim at the post-processing of the images compressed by the HEVC~\cite{sullivan2012overview} intra mode, \eg, Dai~\etal~\cite{dai2017convolutional}, DS-CNN-I \cite{yang2017decoder} and QE-CNN-I \cite{yang2018enhancing}. Most recently, Xing~\etal~\cite{xing2020early} proposed a dynamic DNN to blindly enhance the images compressed with different quality, and showed the effectiveness on both JPEG- and HEVC-compressed images.

Besides, there is more and more interest in training end-to-end DNNs for learned image compression~\cite{Toderici2016Variable,toderici2017full,agustsson2017soft,theis2017lossy,balle2017end,balle2018variational,minnen2018joint,mentzer2018conditional,li2018learning,johnston2018improved,lee2019context,Hu2020Coarse}. For instance, a compressive auto-encoder is proposed in \cite{theis2017lossy}, which achieves comparable performance with JPEG 2000~\cite{skodras2001jpeg}. Then, in \cite{balle2017end} and \cite{ball2018variational}, Balle~\etal~proposed jointly training the auto-encoder with the factorized and hyperpior entropy model, respectively. Meanwhile, Fabian~\etal~\cite{mentzer2018conditional} adopted a 3D-CNN to learn the conditional probability of the elements in latent representations. Later, the hierarchical prior~\cite{minnen2018joint} and the context adaptive~\cite{lee2019context} entropy models were proposed, and successfully outperform the latest image compression standard BPG~\cite{BPG}. Most recently, the coarse-to-fine entropy model was proposed in \cite{Hu2020Coarse} to fully explore the spatial redundancy and achieves the state-of-the-art learned image compression performance. Moreover, several approaches~\cite{Toderici2016Variable,toderici2017full,johnston2018improved} proposed recurrently encoding the residual to compress images at various bit-rates with a single learned model.

However, all aforementioned approaches utilize the trained DNNs at the decoder side, and therefore they are incompatible with the standard image decoders which are widely used in personal computers and mobiles. This limits their applicability in practical scenarios. To overcome this shortcoming, this paper proposes improving rate-distortion performance without modifying the standard JPEG decoder. As far as we know, Talebi~\etal~\cite{talebi2020better} is the only work on pre-editing before JPEG compression, which trains a DNN in the pixel domain before the JPEG encoder to pre-edit input images. Different from \cite{talebi2020better}, we propose learning to improve the  JPEG encoder in the frequency domain, \ie, learning an attention map to apply spatial weighting to the DCT coefficients and learning the quantization tables to optimize rate-distortion performance.

\section{The Proposed Approach}

\subsection{The JPEG Algorithm}
We first briefly introduce the JPEG algorithm.
The first step in JPEG compression is to convert the input image from the RGB color space to the YCbCr colorspace. Next, the image is divided into blocks of $8 \times 8$ pixels which we index by $(n,m) \in [1,N]\times[1,M] $. Each block is then transformed through the (forward) discrete cosine transform ((F)DCT) into frequency space. We denote the DCT coefficients of block $(n,m)$ for the luminance channel $Y$ with  $\ten F^{(Y)}[n,m] \in \mathbb{R}^{8 \times 8}$ and accordingly for channels $Cb, Cr$. Subsequently, the DCT coefficients are quantized using two quantization tables: $\mat Q^{(L)}$ for the luminance channel $Y$ and $\mat Q^{(C)}$ for the chrominance channels $Cb,Cr$. Quantization is applied through elementwise division by the quantization table followed by the rounding function, i.e. for a block in the $Y$ channel: 
\begin{equation}
   \Hat{ Z}_{u,v}^{(Y)} = \Bigg \lfloor \dfrac{ F_{u,v}^{(Y)}} {Q_{u,v}^{(L)}} \Bigg \rceil \text{ for } u,v \in [1,8] 
\end{equation}
Finally, the quantized DCT coefficients are encoded by lossless entropy coding resulting in the compressed image file that also stores the quantization tables. 

\begin{figure}[!t]
\centering
\includegraphics[width=\linewidth]{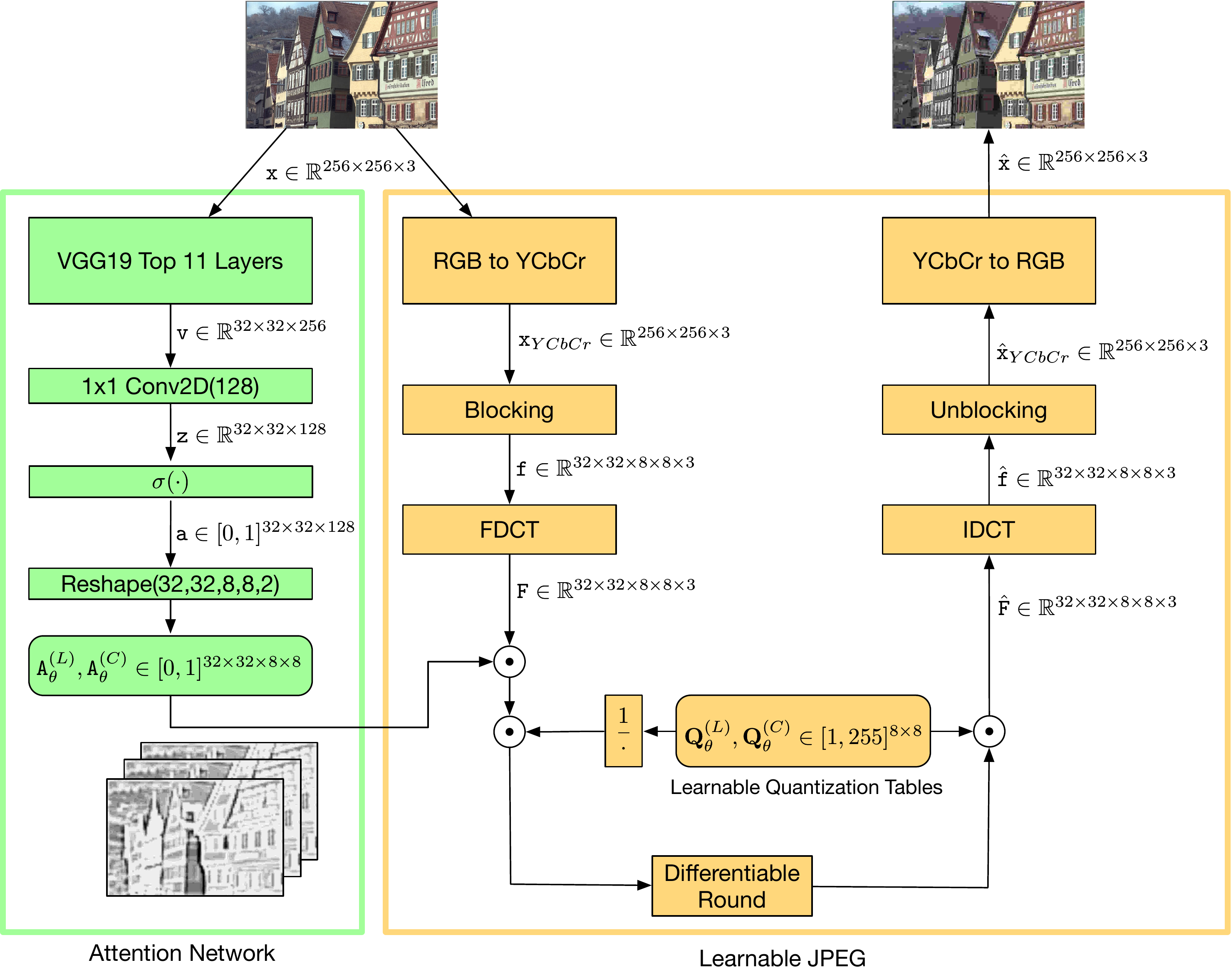}
\caption{Network architecture of the proposed solution}
\label{fig:net}
\end{figure}

\subsection{Proposed Network Architecture} 

In \myfigref{fig:net}, we show a graphical representation summarizing the input-output flow of our proposed architecture. As \myfigref{fig:net} shows, we propose an attention network to pre-edit the input image in the frequency domain by weighting the DCT coefficients, and propose learning the quantization table jointly with the attention network.
To optimize our DNN-based approach on the JPEG encoder, 
we adopt the differentiable JPEG encoder-decoder pipeline as introduced by Shin \etal ~\cite{Shin2017JPEGresistantAI}. In particular this pipeline leaves out the non-differentiable entropy coding and decoding because it is lossless and does not impact the reconstruction loss. Additionally, the rounding operation in the quantization step is replaced by a differentiable 3rd order approximation:
 \begin{equation}
\begin{gathered}
\lfloor x \rceil_{approx} =   \lfloor x \rceil + (\lfloor x \rceil -x)^3.
\end{gathered}
\end{equation}
In our implementation we use this approximation to backpropagate the gradient and use true rounding in forward evaluation.
We can summarize our differentiable encoder-decoder architecture by defining the encoder and decoder functions as:
\begin{equation}
     \Hat{\ten Z} = E_\theta( \ten x), \quad \hat{\ten x} = D_\theta(\Hat{\ten Z}) \implies \quad \hat{\ten x} = D_\theta(E_\theta( \ten x) ).
     \label{autoenc}
\end{equation}
where $\ten x$ is the RGB input image, $\hat {\ten x}$ is the reconstructed RGB output image and $\Hat{\ten Z}$ are the quantized DCT coefficients. 

\textbf{Image Editing through Attention.}
We propose a novel approach to pre-editing the image before quantization to improve the compression quality. \cite{talebi2020better} has shown that an image smoothing network before compressing the image improves the compression performance. We also employ a smoothing mechanism that acts on the DCT coefficients directly. We use a parallel branch in the architecture shown in \myfigref{fig:net} that extracts image features from the original image using a pre-trained version of VGG-19, which is a 19 layer variant of the VGG network \cite{simonyan2014deep}. In particular we use the output of the third $2 \times 2$ max-pooling layer that has $256$ channels. Note that after three max-pooling operations the spatial resolution of the image is reduced by a factor of $8$. Recalling that each $8 \times 8$ block after the FDCT is represented by $64$ DCT coefficients, we can just interpret the DCT coefficients as subchannels to each $Y, Cr, Cb$ channel and get a feature map with the same spatial dimensions as the third layer VGG feature map. We then use a $1 \times 1$ convolutional layer to reduce the channel dimension to $128$.  By using a sigmoid activation we limit the outputs to the range $[0,1] \subset \mathbb{R}$. Now we reshape the $128$ output channels to a $8 \times 8 \times 2$ tensor for each block. We split this tensor into 2 giving us the attention tensors for luminance and chrominance for all $N \cdot M$ blocks:
\begin{equation}
\begin{gathered}
\ten A^{(L)} \in \{x \in \mathbb{R} \mid 0 \leq x \leq 1\}^{N \times M \times 8 \times 8}, \\
\ten A^{(C)} \in \{x \in \mathbb{R} \mid 0 \leq x \leq 1\}^{N \times M \times 8 \times 8}.
\end{gathered}
\end{equation}
We now multiply each DCT coefficient by its importance score in \myeqref{quantize} before we apply the learnable quantization table.

\textbf{Learnable quantization table.}
In the following, we 
use the differentiable JPEG pipeline to learn the quantization tables by introducing $\mat Q_\theta^{(L)}$ and $ \mat Q_\theta^{(C)}$ as optimization variables. We use the subscript $\theta$ to indicate that this quantity is learned. We limit the value range of the quantization tables to $[1,255]$ by specifying a clipping function that adjusts the range after an optimization step.

Using the Hadamard product $\odot$ for ease of notation, the proposed approach can be summarized as:
\begin{equation}
\begin{gathered}
\Hat{\ten Z}^{(Y)}[n,m] = \Bigg \lfloor \ten F^{(Y)}[n,m]  \odot \ten A^{(L)}[n, m]  \odot \ {\Bar{\mat Q}_\theta^{(L)}}\Bigg \rceil_{approx}\\
\Hat{\ten Z}^{(Cr)}[n,m] = \Bigg \lfloor \ten F^{(Cr)}[n,m]  \odot \ten A^{(C)}[n, m] \odot \ {\Bar{\mat Q}_\theta^{(C)}}\Bigg \rceil_{approx}\\
  \Hat{\ten Z}^{(Cb)}[n,m] = \Bigg \lfloor \ten F^{(Cb)}[n,m]  \odot \ten A^{(C)}[n, m]  \odot \ {\Bar{\mat Q}_\theta^{(C)}}\Bigg \rceil_{approx}\\
   \quad \text{ for } n \in [1,N], m \in [1,M],\\
   \text{with }
 \Bar{Q}_{u,v}^{(L)} = \dfrac{1} {Q_{u,v}^{(L)}}, \quad
  \Bar{Q}_{u,v}^{(C)} = \dfrac{1} {Q_{u,v}^{(C)}},
  \quad \text{ for } u,v \in [1,8].
\end{gathered}
\label{quantize}
\end{equation}
It is important to understand that this modification is not recoverable in the decoder. Multiplying the DCT coefficients by a number smaller or equal to 1 acts like a frequency filter. Typically, higher frequencies get suppressed more so we get a low-pass filter. By distributing the attention weights across the spatial dimension and the DCT-coefficient dimension we get a smoothing filter that is adaptive spatially and across different frequencies. The big advantage of using such an attention mechanism instead of a feedforward smoothing network is that we can control the flow of information by limiting the norm of the attention maps.
Note that, to get the final JPEG image for evaluation, we perform  the entropy coding step with the default Huffman tables used for JPEG.

\subsection{Evaluation Metrics}

\noindent A widely used metric for measuring image similarity is the Mean Squared Error (MSE) that can be converted to the Peak Signal to Noise Ratio (PSNR) on a logarithmic scale. Similarly to \cite{Cavigelli_2017} we define the MSE and PSNR for the tensors $\ten x, \hat{ \ten x} \in \mathcal{X}$ of arbitrary dimension:
\begin{equation}
\begin{gathered}
    \text{MSE}( \ten x, \hat{ \ten x}) = \dfrac{1}{\vert\mathcal{P}\vert}\sum_{\vec p \in \mathcal{P}}( x_{\vec p} - \hat{ x}_{\vec p})^2\\
    \text{PSNR}(\ten x, \hat{ \ten x}) = 10 \log_{10} \left( \dfrac{255^2}{\text{MSE}( \ten x, \hat{ \ten x})} \right)\\
    \text{where } \mathcal{P} \text{ is the set of pixel indices and } x_{\vec p}, \hat{x}_{\vec p} \in [0,255], \quad \forall \vec p \in \mathcal{P}.
\end{gathered}
\label{MSE}
\end{equation}
To better represent local statistics we also use the Multi-Scale Structural Similarity (MS-SSIM) \cite{mssim} with its default implementation \textit{tf.image.ssim\_multiscale} in Tensorflow.
With deep neural networks dominating computer vision tasks, Zhang~\etal~\cite{zhang2018unreasonable} have developed the Learned Perceptual Image Patch Similarity (LPIPS) metric that leverages the power of deep features to judge image similarity. We use the version of this metric that is based on AlexNet  \cite{Krizhevsky_imagenetclassification} features.
For measuring the strength of the compression we define the bit rate in bits per pixel as:
\begin{equation}
    \text{BPP} \left[ \dfrac{\text{bit}}{\text{px}} \right] = \dfrac{\text{file size }\text{[bit]}}{\text{total number of pixels }\text{[px]}}.
\end{equation}

 \subsection{Formulation of the Loss}
 Optimizing a compression task has to tackle the  fundamental balance between quality and file size that we introduced as the rate-distortion tradeoff. 
  For the RGB input image $\ten x$ and the reconstructed RGB image $\hat {\ten x}$, given the learned parameters $\theta$, our loss function has the general form:
\begin{equation}
 \begin{gathered}
 \mathcal{L}(\ten x, \hat {\ten x}; \theta) = \lambda \cdot \text{d}(\ten x, \hat {\ten x}) + \text{r}(\ten x, \hat {\ten x}; \theta),\\
 \text{with }  \ten x, \hat {\ten x} \in \{ t \in \mathbb{R} \mid 0 \leq t \leq 255\}^{8N \times 8M  \times 3}, \quad \lambda \in \mathbb{R}^+,
 \end{gathered}
 \end{equation}
 where we call $\text{r}(\ten x, \hat {\ten x}; \theta)$ the rate loss and $\text{d}(\ten x, \hat {\ten x})$ the distortion loss.
\noindent The parameter $\lambda$ determines the ratio of the two major loss terms and hence balances distortion and rate. Usually the network is trained for several values of $\lambda$ to obtain a rate-distortion curve.
 
 \subsubsection{Distortion Loss.}
 The distortion loss measures how close the reconstructed image is to the original image. We use the combination of the fidelity loss MSE \myeqref{MSE} and the perceptual loss LPIPS:
  \begin{equation}
 \text{d}(\ten x, \hat {\ten x}) = \text{MSE}(\ten x, \hat {\ten x}) + \gamma \cdot \text{LPIPS}(\ten x, \hat {\ten x})
  \end{equation}
 where we introduce $\gamma$ as the LPIPS loss weight.
 
\subsubsection{Rate Loss.}
The final bit per pixel rate that determines the file size of the JPEG file is proportional to the discrete entropy of the quantized DCT coefficients, given that the consecutive Huffman encoding is optimal. Because of the previously mentioned differentiability issue, we cannot directly add the entropy as a loss term to be minimized. Instead, \cite{talebi2020better} uses the differentiable entropy estimator proposed in \cite{ball2016endtoend} that relies on a continuous density estimator explained in the appendix of \cite{ball2018variational}. Using this estimator to optimize our learned JPEG architecture did not result in improved performance over the standard JPEG. We hence propose a novel formulation of the rate loss that is entirely regularization based:
\begin{equation}
 \text{r}(\ten x; \theta) = \alpha(\Vert\bar{\mat Q}_\theta^{(L)}\Vert_1 + \Vert\bar{\mat Q}_\theta^{(C)}\Vert_1) + \beta (\text{mean}(\ten A_\theta^{(L)}(\ten x)) + \text{mean}(\ten A_\theta^{(C)}(\ten x)))
 \label{rate_loss}
  \end{equation}
with the mean function:
\begin{equation}
\begin{gathered}
    \text{mean}( \ten A) = \dfrac{1}{\vert\mathcal{P}\vert}\sum_{\vec p \in \mathcal{P}} A_{\vec p},\\
\end{gathered}
\label{MSE}
\end{equation}
where $\mathcal{P}$ is the index set over all entries in the tensor $\ten A$.
The intuition behind this is that large quantization values reduce the entropy monotonically. It is important to note that these loss terms are only related to the reduction in entropy and not the entropy itself. For optimization with gradient descent, a constant offset of the true entropy does not affect the gradient as it vanishes in the derivative. We denote the attention maps as functions of $\ten x$ here to point out that they are input dependent, the quantization tables on the other hand are not a function of the input. 

Based on \myeqref{autoenc} and the loss definitions above we can formulate the optimization objective as:
\begin{equation}
    \underset{\theta}{\min} \,\, \mathcal{L}(\ten x, D_\theta(E_\theta( \ten x) ); \theta).
\end{equation}

\section{Experiments}
\subsection{Datasets}
The learned JPEG network is trained on the dataset provided by Hasinoff~\etal~\cite{hasinoff}. It consists of 3640 HDR image bursts with a resolution of 12 megapixels. For training we use the merged HDR images provided in the dataset. We extract image patches of size $256$ that are obtained from randomly cropping.
We evaluate our model on the Kodak dataset \cite{kodak}, consisting of 24 uncompressed images of size $768 \times 512$. Additionally, we evaluate on the DIV2K \cite{div2k,Timofte_2017_CVPR_Workshops} validation set that contains 100 high quality images with 2040 pixels along the long edge.

\subsection{Training Procedure}
The Tensorflow implementation of our model is trained using gradient descent with the Adam \cite{kingma2014adam} optimizer. The learning rate was set to $10^{-6}$. We use a batch size of $8$ throughout the experiments. During training, the rate distortion parameter $\lambda$ is varied in the interval $[10^{-4}, 10^{-1}]$. For any choice of $\lambda$, the network is trained for 20000 steps. We used GPU based training on the NVIDIA TITAN X as well as the NVIDIA GTX 1080Ti for training in most experiments. For the single image training we used the NVIDIA TESLA P100.

The VGG layers of the attention network are initialized using weights from pretraining on ImageNet \cite{imagenet_cvpr09} and refined during training. The  $1 \times 1$ convolutional layer is initialized with the Glorot uniform initializer. The quantization tables optimization variables are initialized uniformly in the interval $[1s,2s]$ and are limited to the range $[1s, 255s]$ where $s > 0$ is a scaling factor, which we set to $s=10^{-5}$. The final quantization tables are calculated by multiplying with $s^{-1}$. This choice was made because the quantization tables are trained jointly with the neural network weights which generally have a much smaller magnitude. The introduced scaling factor makes up for that difference in magnitude and allows all parameters to be updated with the same optimizer and learning rate. Alternatively, one could use a separate optimizer with an adjusted learning rate to only train the quantization tables.

\begin{figure}[!t]
\centering
\captionsetup[subfigure]{labelformat=empty}
\begin{subfigure}[b]{.32\linewidth}
\includegraphics[width=\linewidth, trim={5 0 40 10}, clip=true]{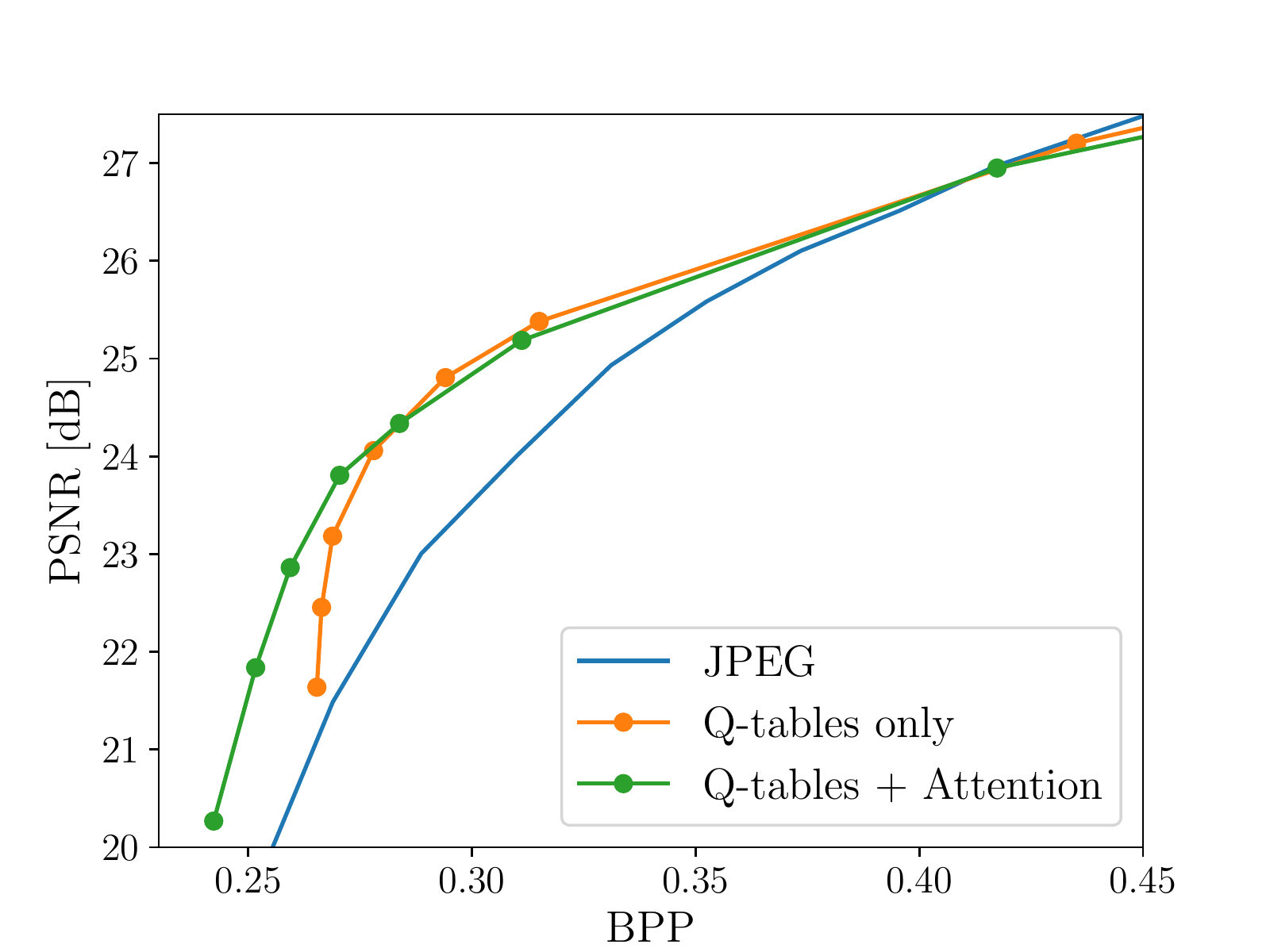}
\caption{PSNR on Kodak}
\end{subfigure}
\hfil
\begin{subfigure}[b]{.32\linewidth}
\includegraphics[width=\linewidth, trim={5 0 40 20}, clip=true]{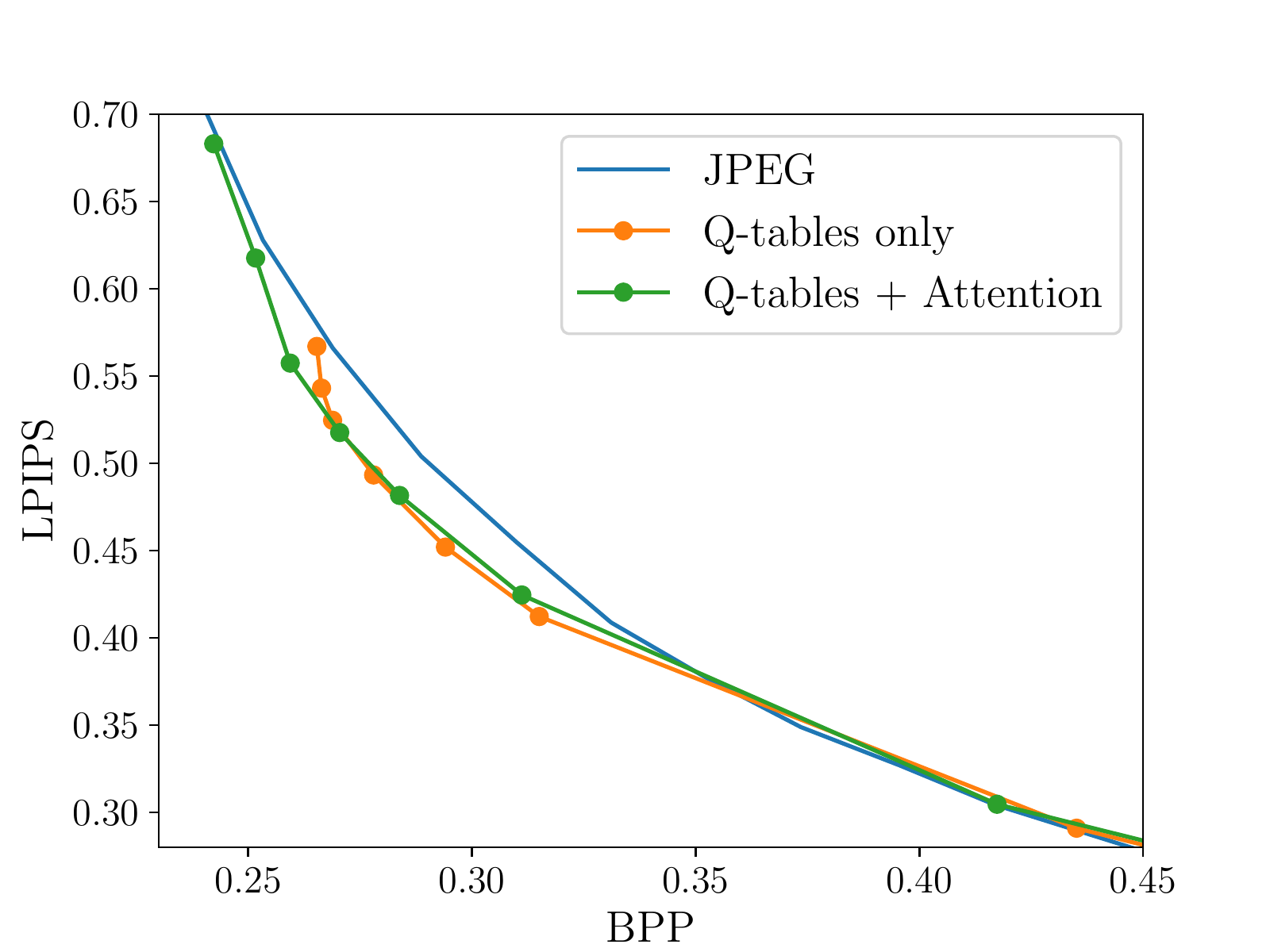}
\caption{LPIPS on Kodak}
\end{subfigure}
\hfil
\begin{subfigure}[b]{.32\linewidth}
\includegraphics[width=\linewidth, trim={5 0 40 20}, clip=true]{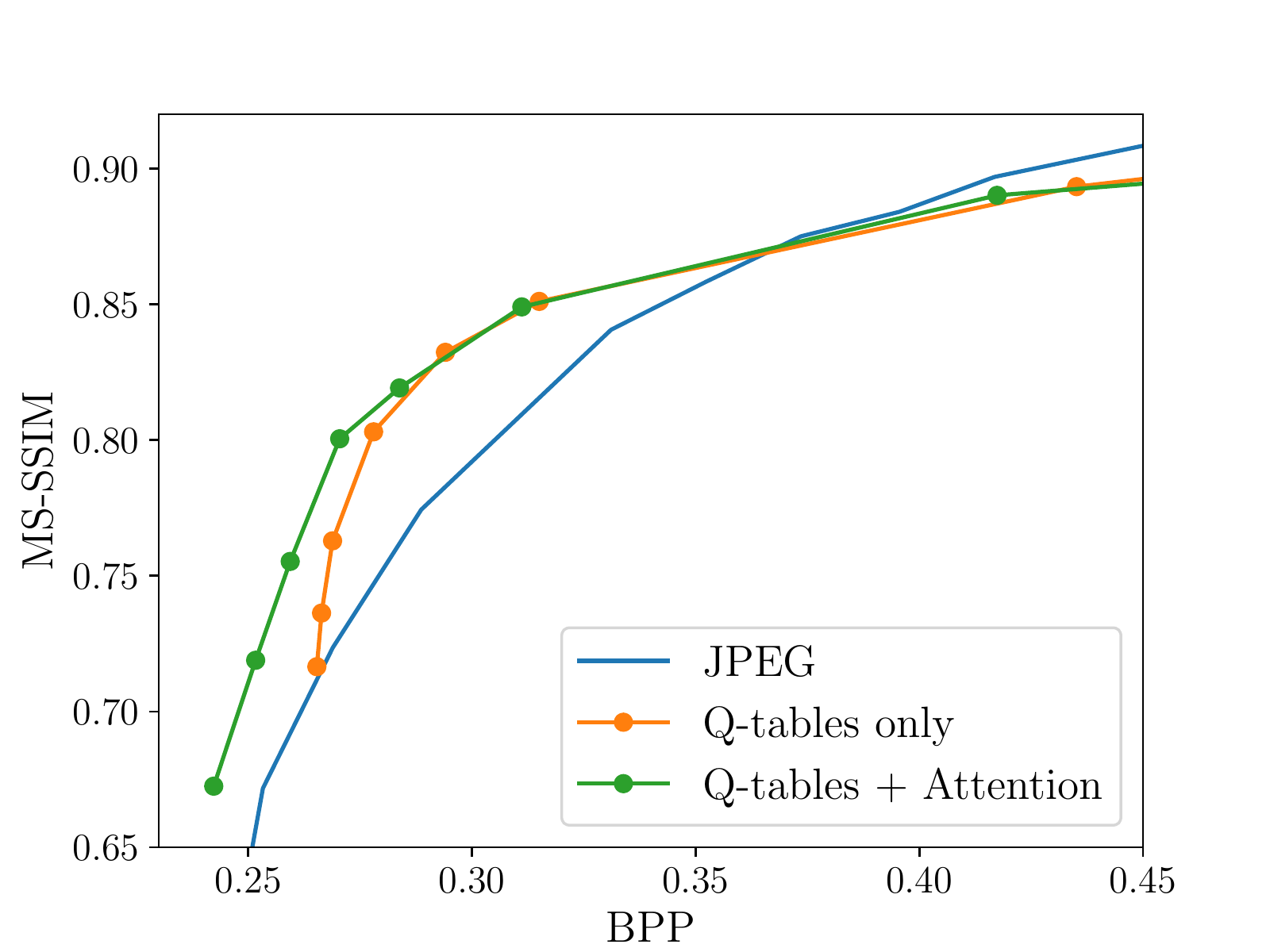}
\caption{MS-SSIM on Kodak}
\end{subfigure}
\begin{subfigure}[b]{.32\linewidth}
\includegraphics[width=\linewidth, trim={5 0 40 20}, clip=true]{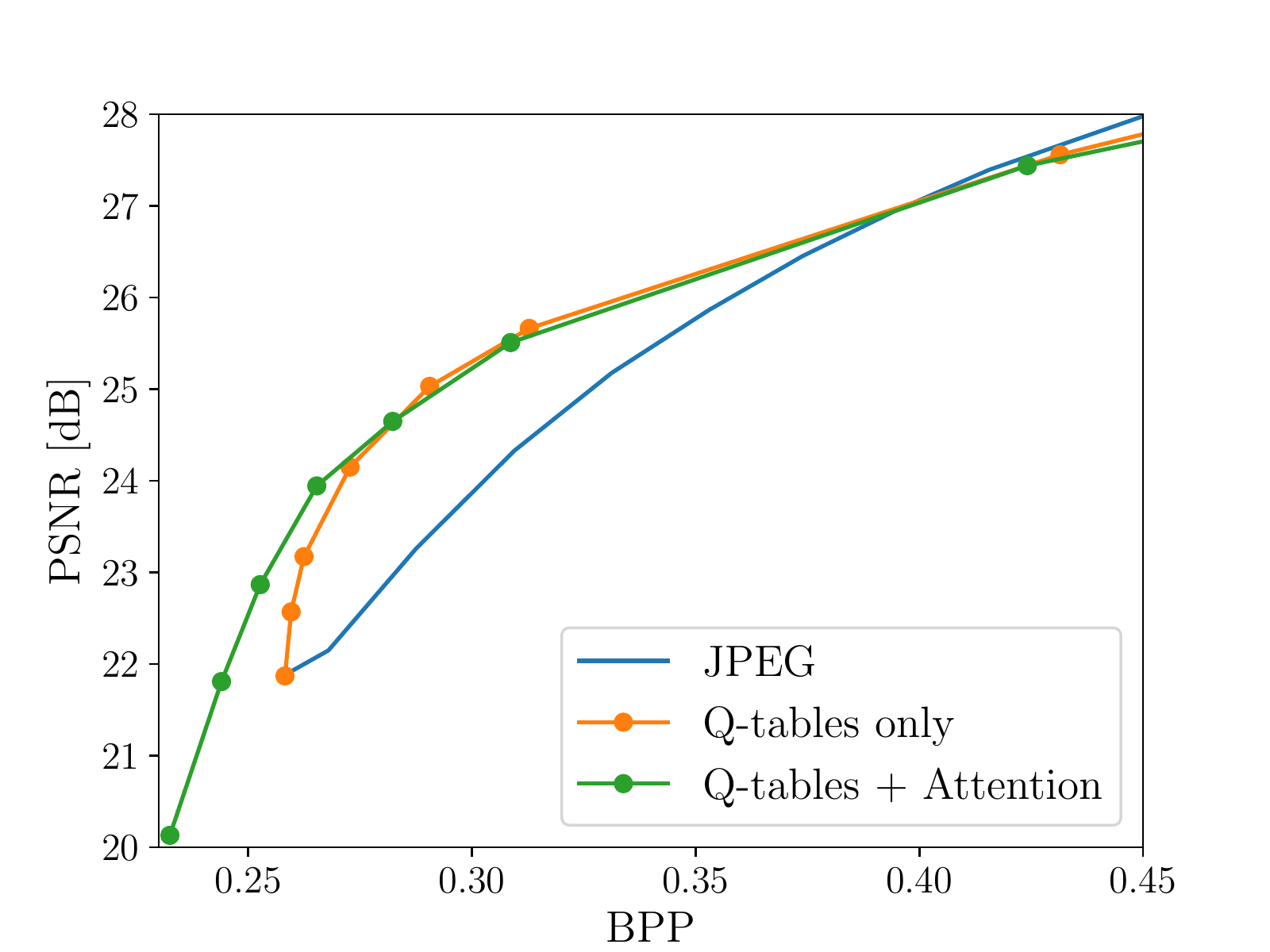}
\caption{PSNR on DIV2K}
\end{subfigure}
\hfil
\begin{subfigure}[b]{.32\linewidth}
\includegraphics[width=\linewidth, trim={5 0 40 20}, clip=true]{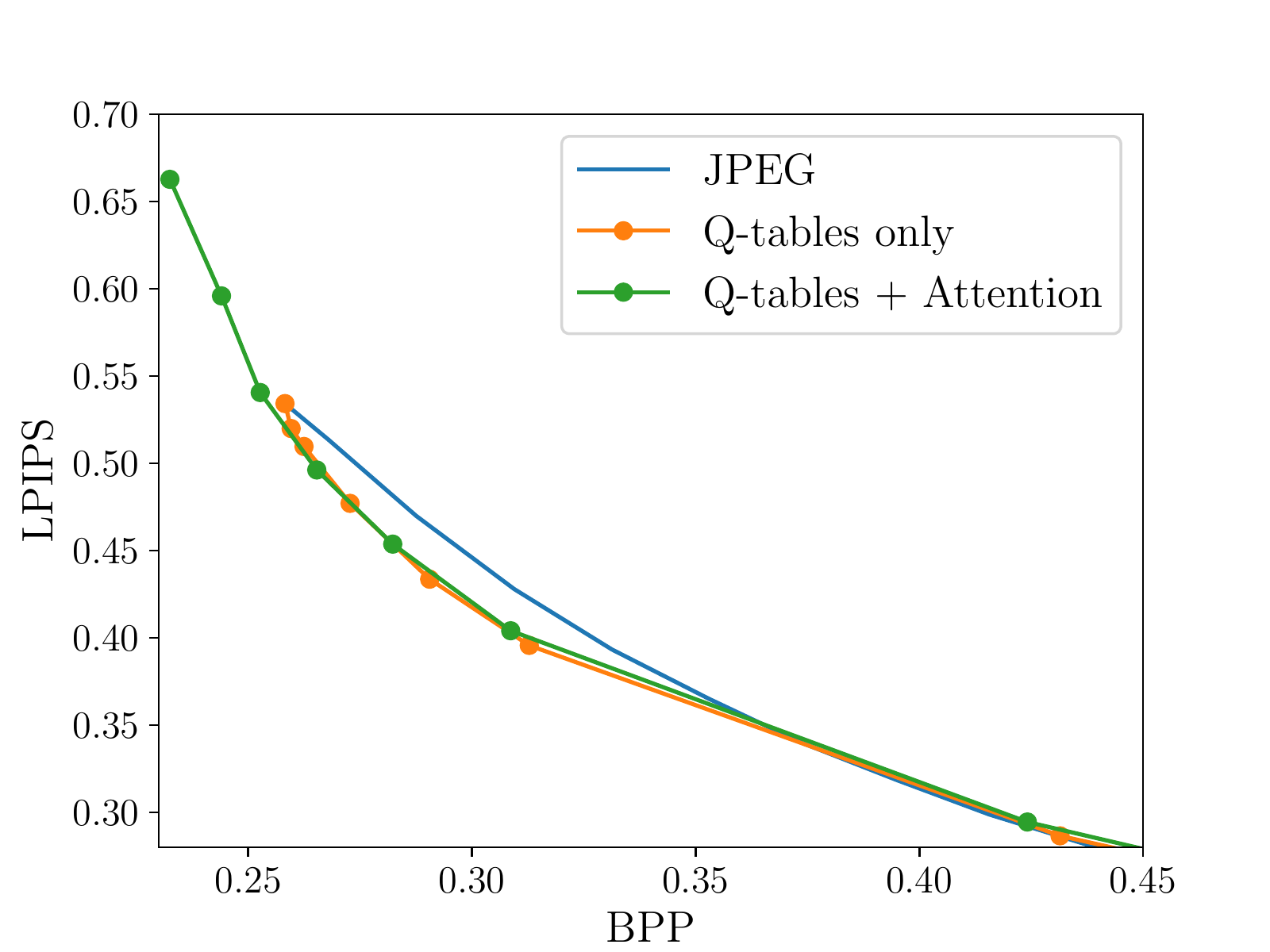}
\caption{LPIPS on DIV2K}
\end{subfigure}
\hfil
\begin{subfigure}[b]{.32\linewidth}
\includegraphics[width=\linewidth, trim={5 0 40 20}, clip=true]{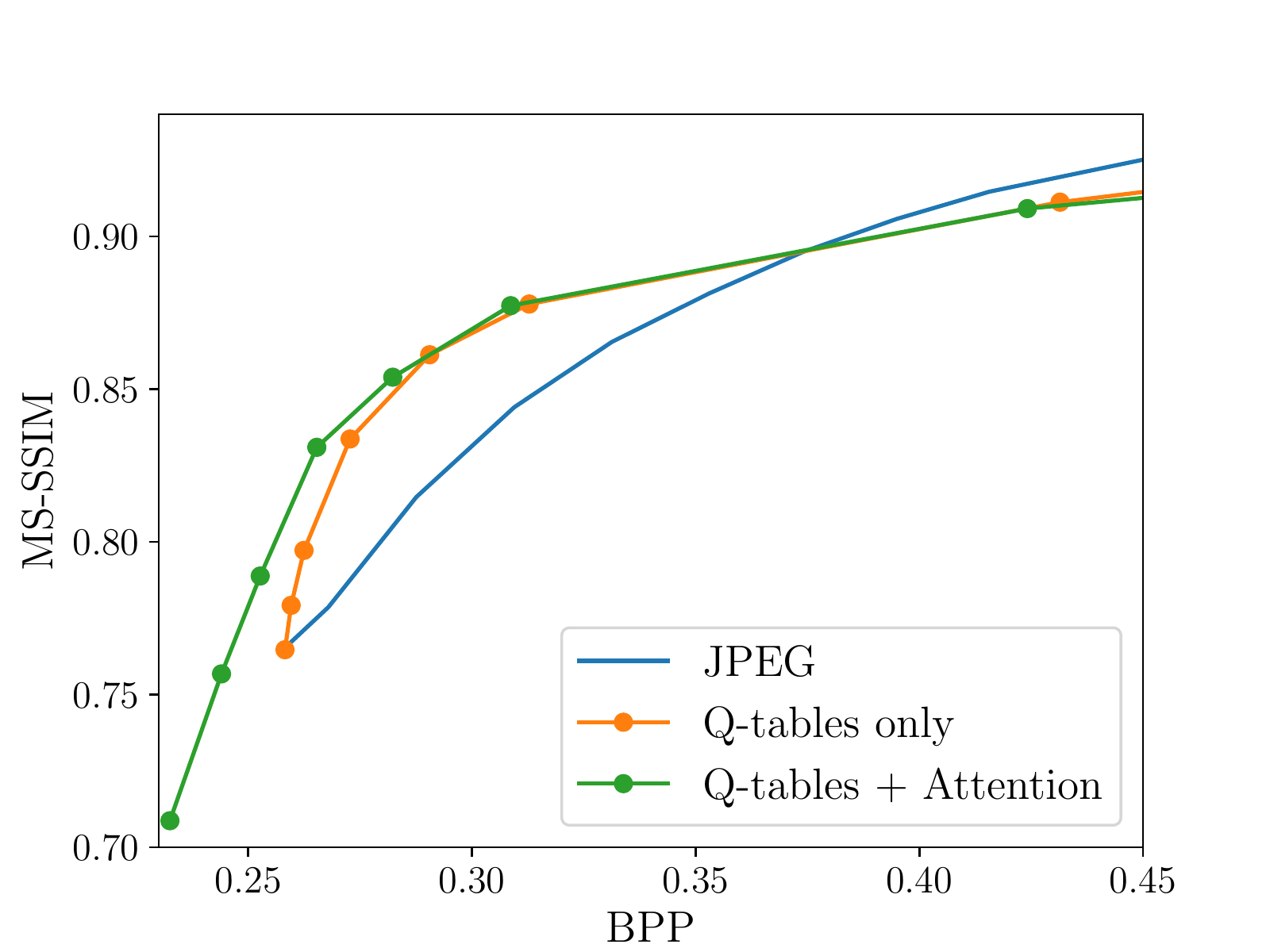}
\caption{MS-SSIM on DIV2K}
\end{subfigure}

\caption{Metrics comparison for learned quantization tables (Q-tables only) against learned  quantization tables together with attention (Q-tables + Attention) and the JPEG baseline.}
\label{fig:attention_only_metric}
\end{figure}

\subsection{Comparison of the Proposed Solution to JPEG}
We use the Python package Pillow to interface the standard JPEG encoder libjpeg. We compute the JPEG baseline by compressing the test images for quality factors in $[1,90]$. We do not use any chroma subsampling and use default Huffman tables for entropy coding.

It is important to note that the plots used for comparison are averaged as follows: The model trained for a certain rate-distortion parameter $\lambda$ is evaluated on the whole test set. We average the bpp as well as the results for each metric over the test set. This gives us one data point per $\lambda$ which we present in the plot.
In \myfigref{fig:attention_only_metric}, we show the performance of learning the quantization tables with and without the attention based editing. We set the hyperparameters to $\alpha=10$, $\beta=1$, $\gamma=500$ which achieve the best results in our experiments.

\begin{figure}[!t]
\centering
\captionsetup[subfigure]{labelformat=empty, font=small}
\begin{subfigure}[b]{.24\linewidth}
\caption{original}
\includegraphics[width=\linewidth]{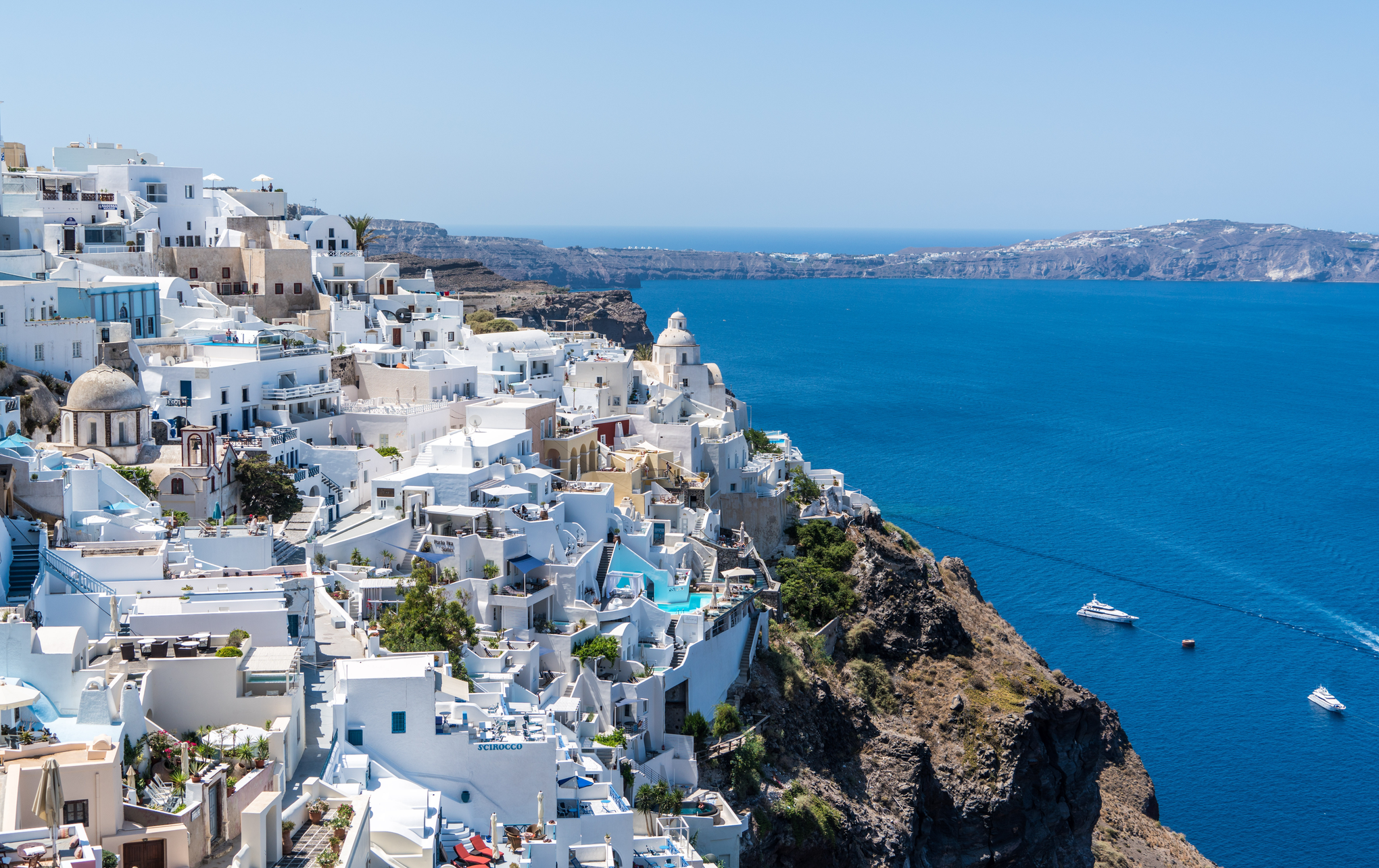}
\caption{$12.4$ bpp}
\end{subfigure}
\hfil
\begin{subfigure}[b]{.24\linewidth}
\caption{JPEG}
\includegraphics[width=\linewidth]{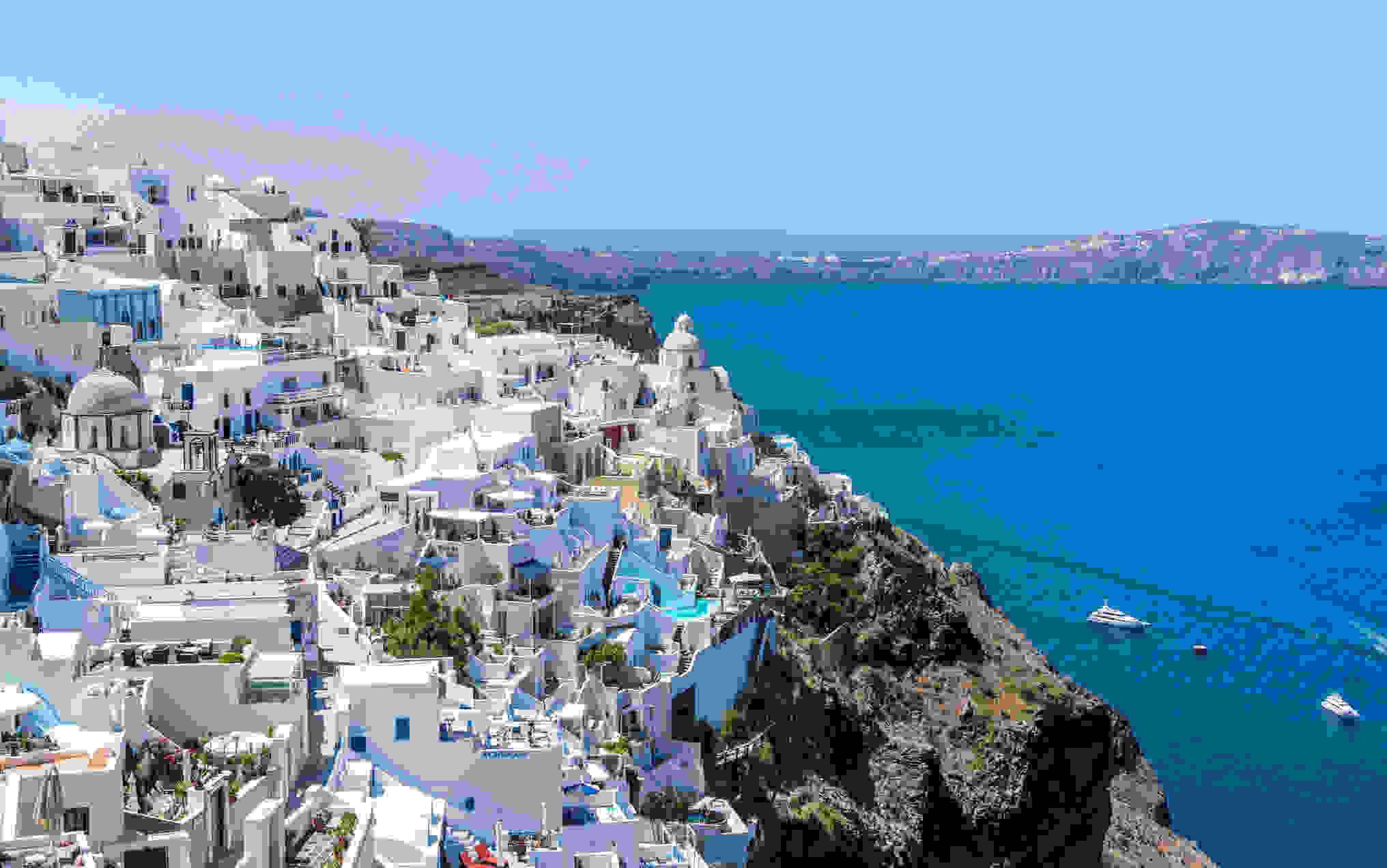}
\caption{$0.270$ bpp}
\end{subfigure}
\hfil
\begin{subfigure}[b]{.24\linewidth}
\caption{Q-tables only}
\includegraphics[width=\linewidth ]{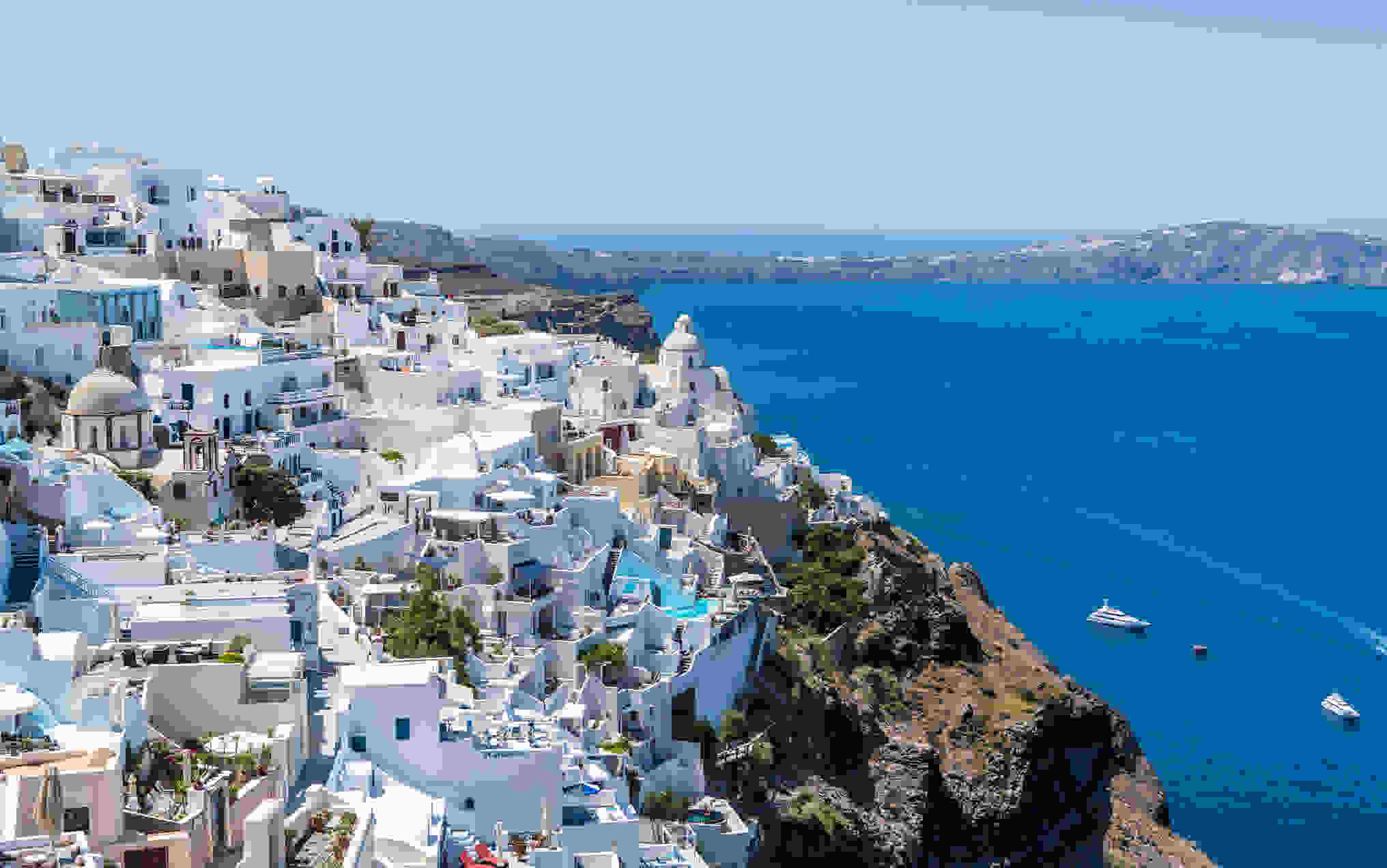}
\caption{$0.263$ bpp}
\end{subfigure}
\hfil
\begin{subfigure}[b]{.24\linewidth}
\caption{Q-tables + Attention}
\includegraphics[width=\linewidth]{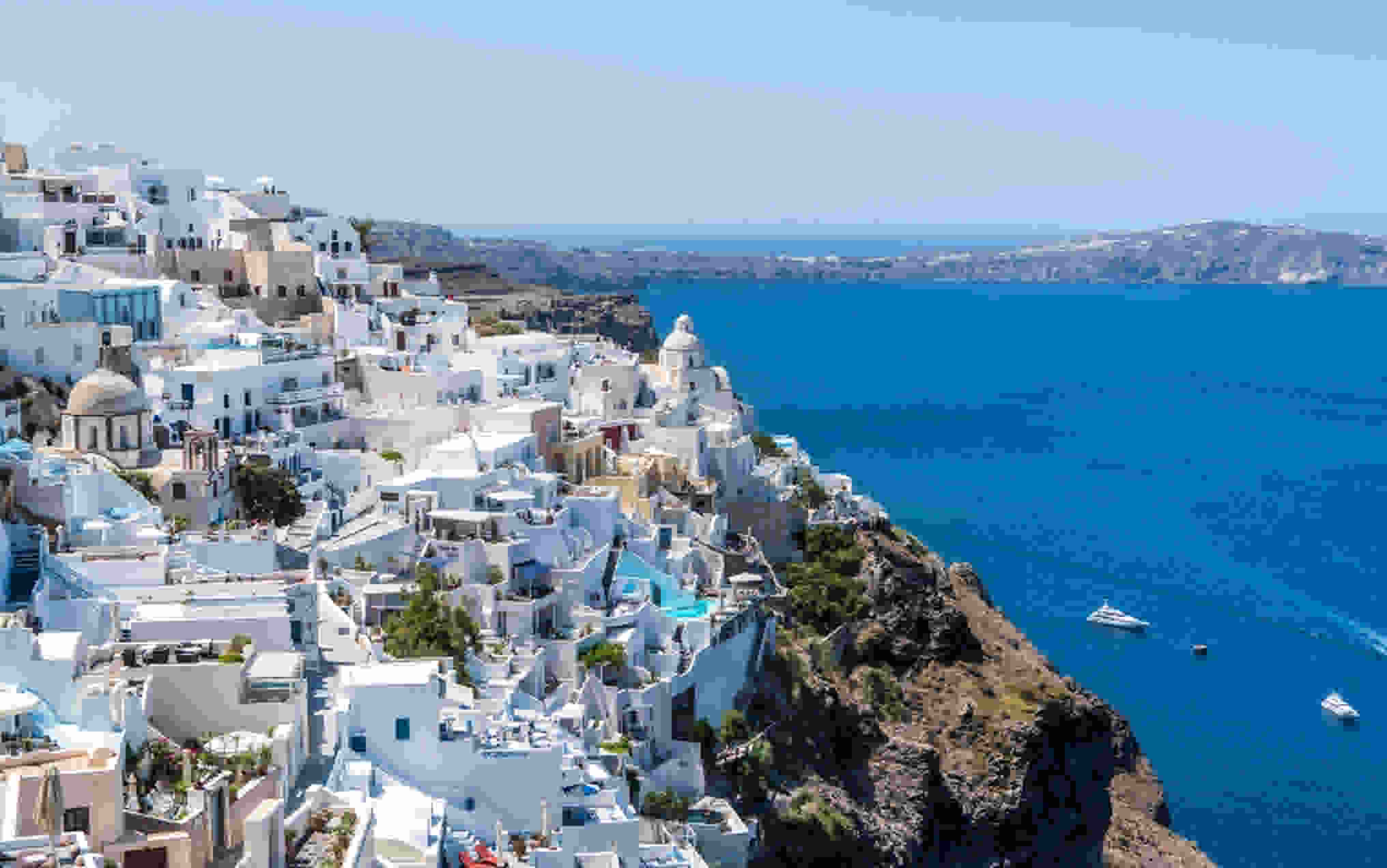}
\caption{$0.264$ bpp}
\end{subfigure}
\begin{subfigure}[b]{.24\linewidth}

\includegraphics[width=\linewidth]{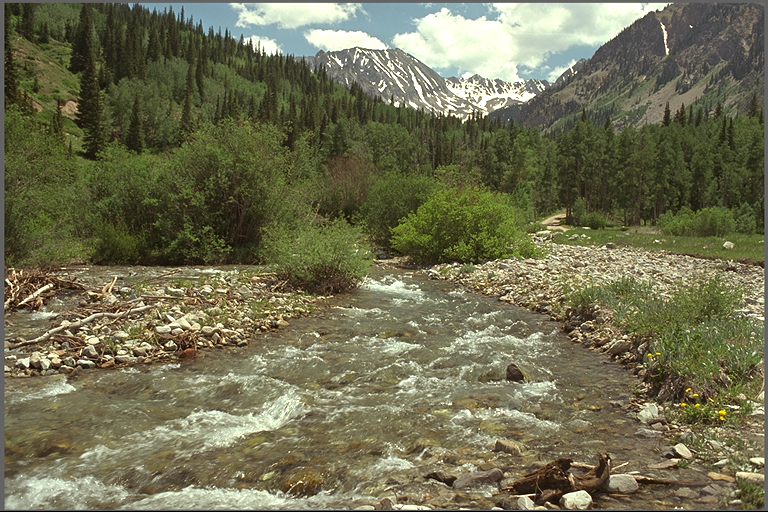}
\caption{$17.2$ bpp}
\end{subfigure}
\hfil
\begin{subfigure}[b]{.24\linewidth}

\includegraphics[width=\linewidth]{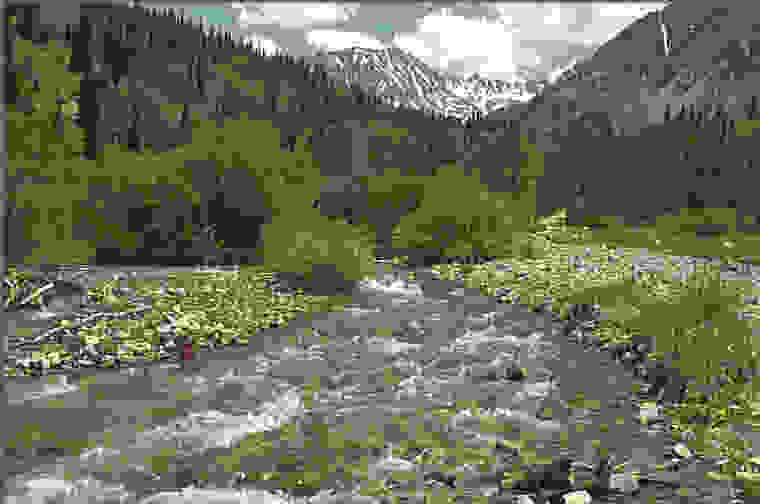}
\caption{$0.297$ bpp}
\end{subfigure}
\hfil
\begin{subfigure}[b]{.24\linewidth}

\includegraphics[width=\linewidth ]{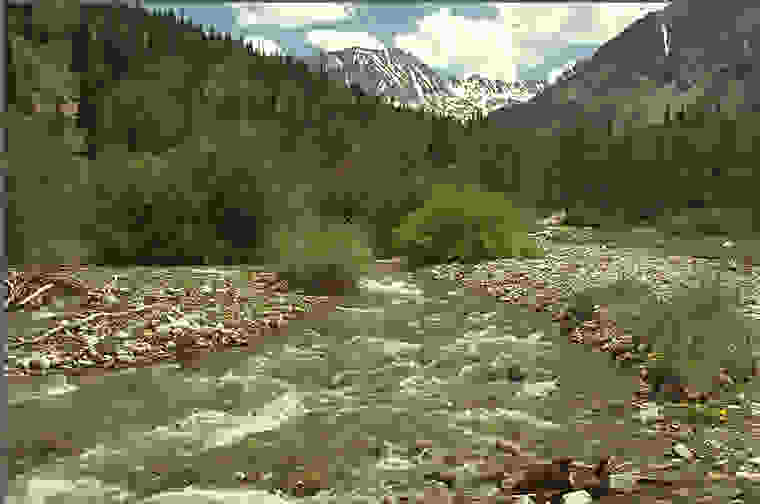}
\caption{$0.297$ bpp}
\end{subfigure}
\hfil
\begin{subfigure}[b]{.24\linewidth}

\includegraphics[width=\linewidth]{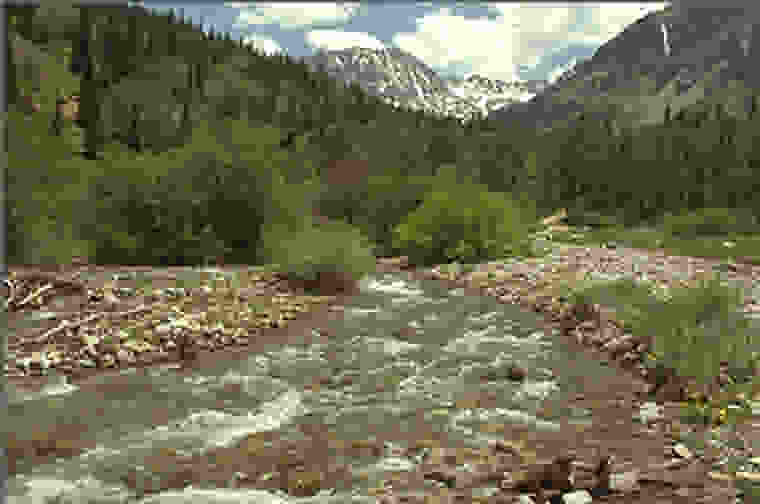}
\caption{$0.292$ bpp}
\end{subfigure}
\begin{subfigure}[b]{.24\linewidth}

\includegraphics[width=\linewidth]{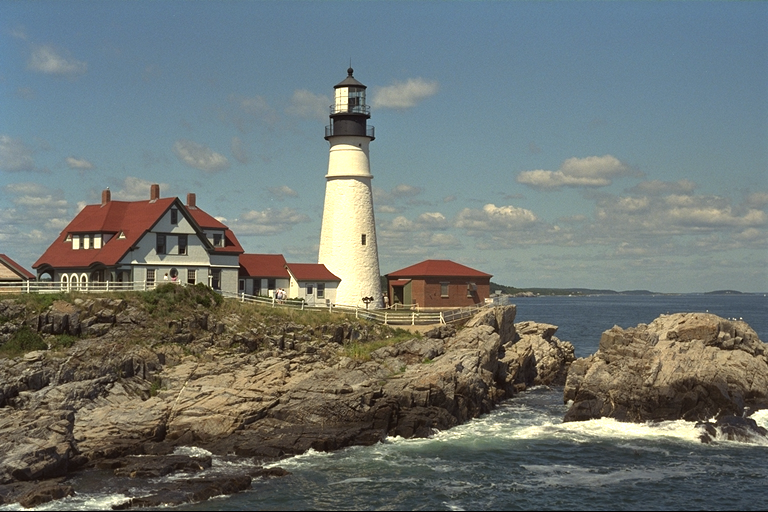}
\caption{$13.3$ bpp}
\end{subfigure}
\hfil
\begin{subfigure}[b]{.24\linewidth}

\includegraphics[width=\linewidth]{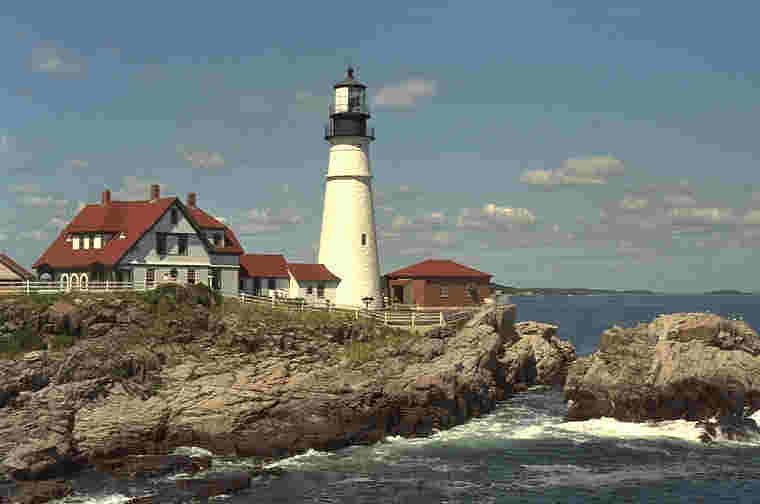}
\caption{$0.444$ bpp}
\end{subfigure}
\hfil
\begin{subfigure}[b]{.24\linewidth}

\includegraphics[width=\linewidth ]{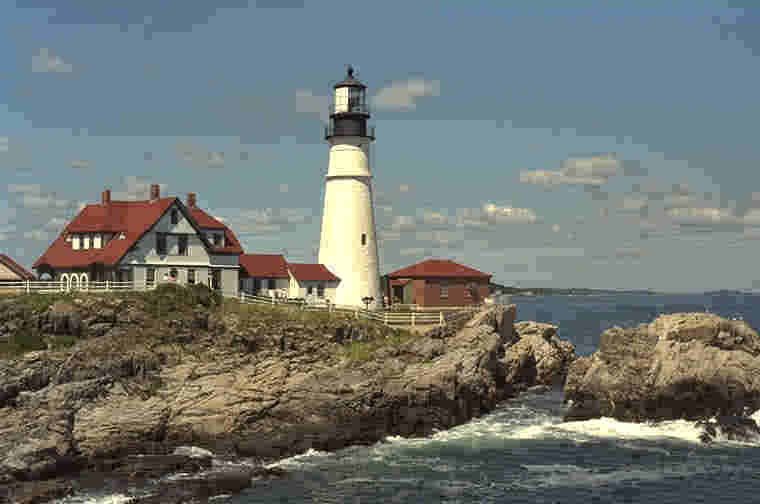}
\caption{$0.443$ bpp}
\end{subfigure}
\hfil
\begin{subfigure}[b]{.24\linewidth}

\includegraphics[width=\linewidth]{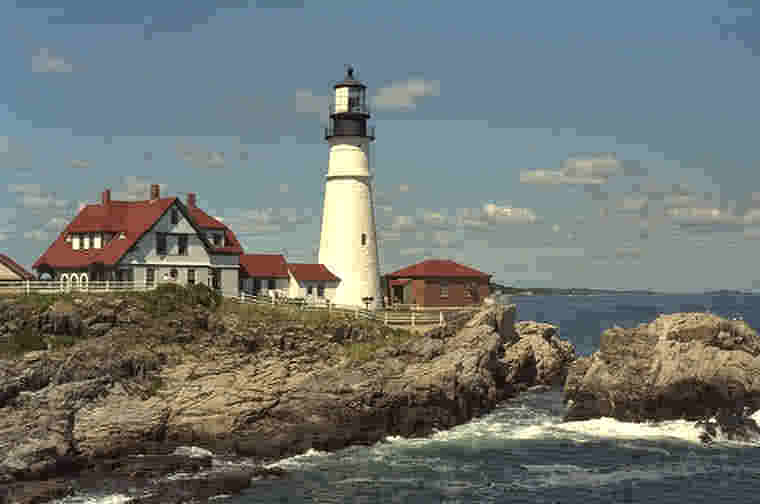}
\caption{$0.424$ bpp}
\end{subfigure}

\begin{subfigure}[b]{.24\linewidth}

\includegraphics[width=\linewidth]{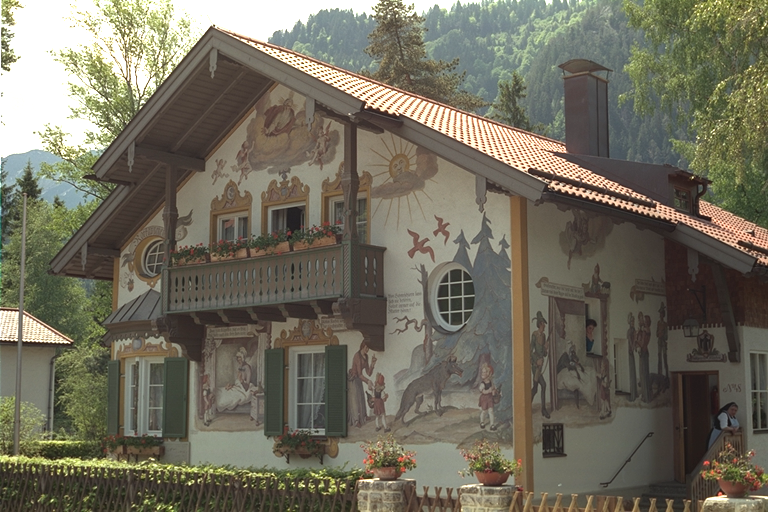}
\caption{$14.8$ bpp}
\end{subfigure}
\hfil
\begin{subfigure}[b]{.24\linewidth}

\includegraphics[width=\linewidth]{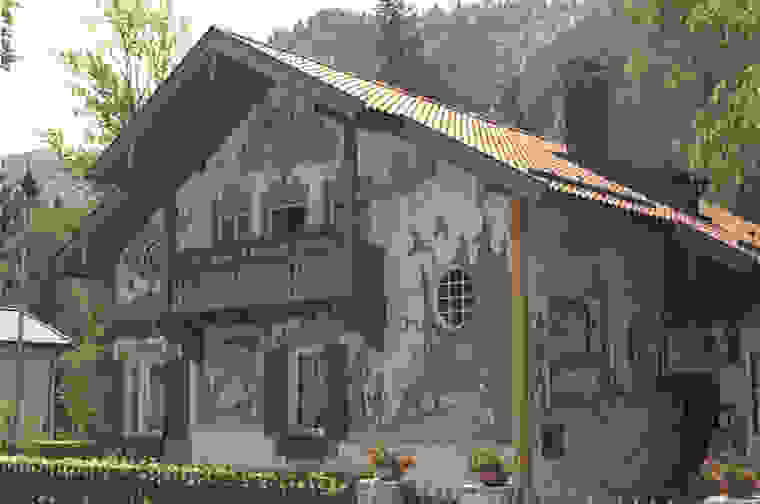}
\caption{$0.278$ bpp}
\end{subfigure}
\hfil
\begin{subfigure}[b]{.24\linewidth}

\includegraphics[width=\linewidth]{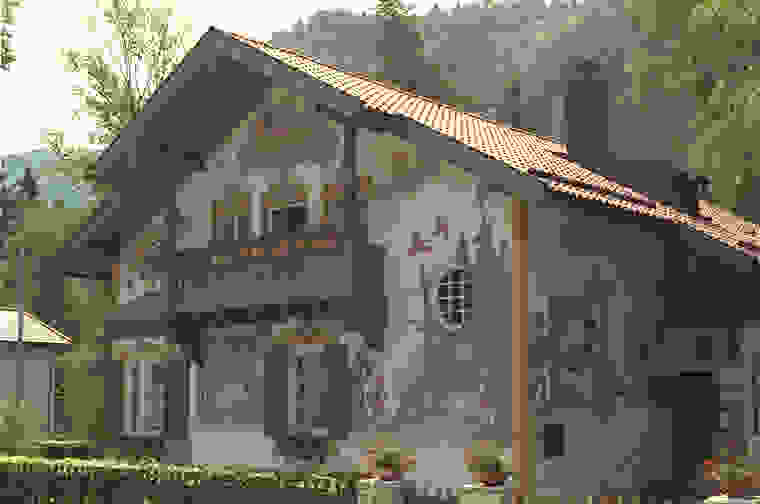}
\caption{$0.278$ bpp}
\end{subfigure}
\hfil
\begin{subfigure}[b]{.24\linewidth}

\includegraphics[width=\linewidth]{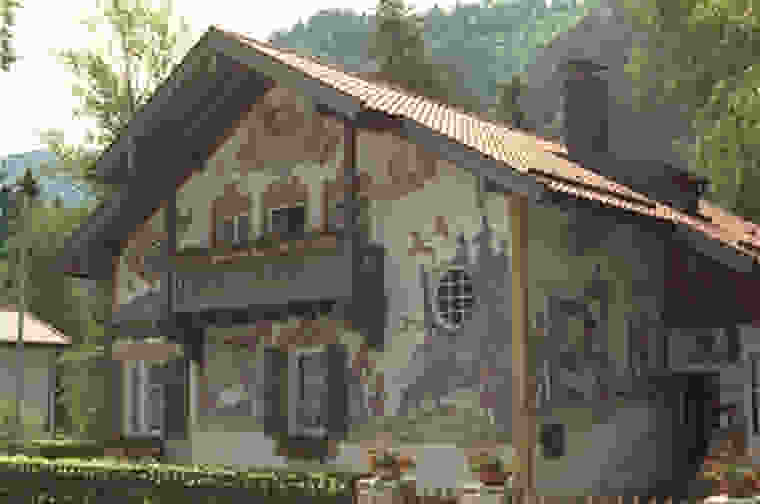}
\caption{$0.279$ bpp}
\end{subfigure}
\caption{Comparison of compressed image output from the JPEG baseline and our proposed solution for learning only the quantization tables (Q-tables only) or learning the quantization tables jointly with the attention map (Q-tables + Attention).}
\label{fig:visual_comp}
\end{figure}

Generally, we see a clear performance increase over the JPEG baseline for both configurations at bpp $< 0.4$. The advantage in PSNR when using the attention network shows in the lowest bit per pixel range where the smoothing allows to further reduce the file size. In \myfigref{fig:visual_comp}, we can see a side-by-side comparison of multiple images compressed with either the standard JPEG encoder, our proposed encoder with trained quantization tables or our proposed encoder with trained quantization tables and attention based editing. The most notable difference is that JPEG performs clearly worse in terms of color accuracy. The addition of the attention network helps retain better color, especially at lower bpp with only slightly sacrificing detail. At higher bpp, \ie, in the third image, the visual quality is similar as without attention but the achieved bitrate is lower.

\subsection{Evaluation of our Rate Loss as a Proxy of BPP}
\begin{figure}[!t]
\centering
\captionsetup[subfigure]{labelformat=empty}
\begin{subfigure}[b]{.32\linewidth}
\includegraphics[width=\linewidth, trim={5 0 40 20}, clip=true]{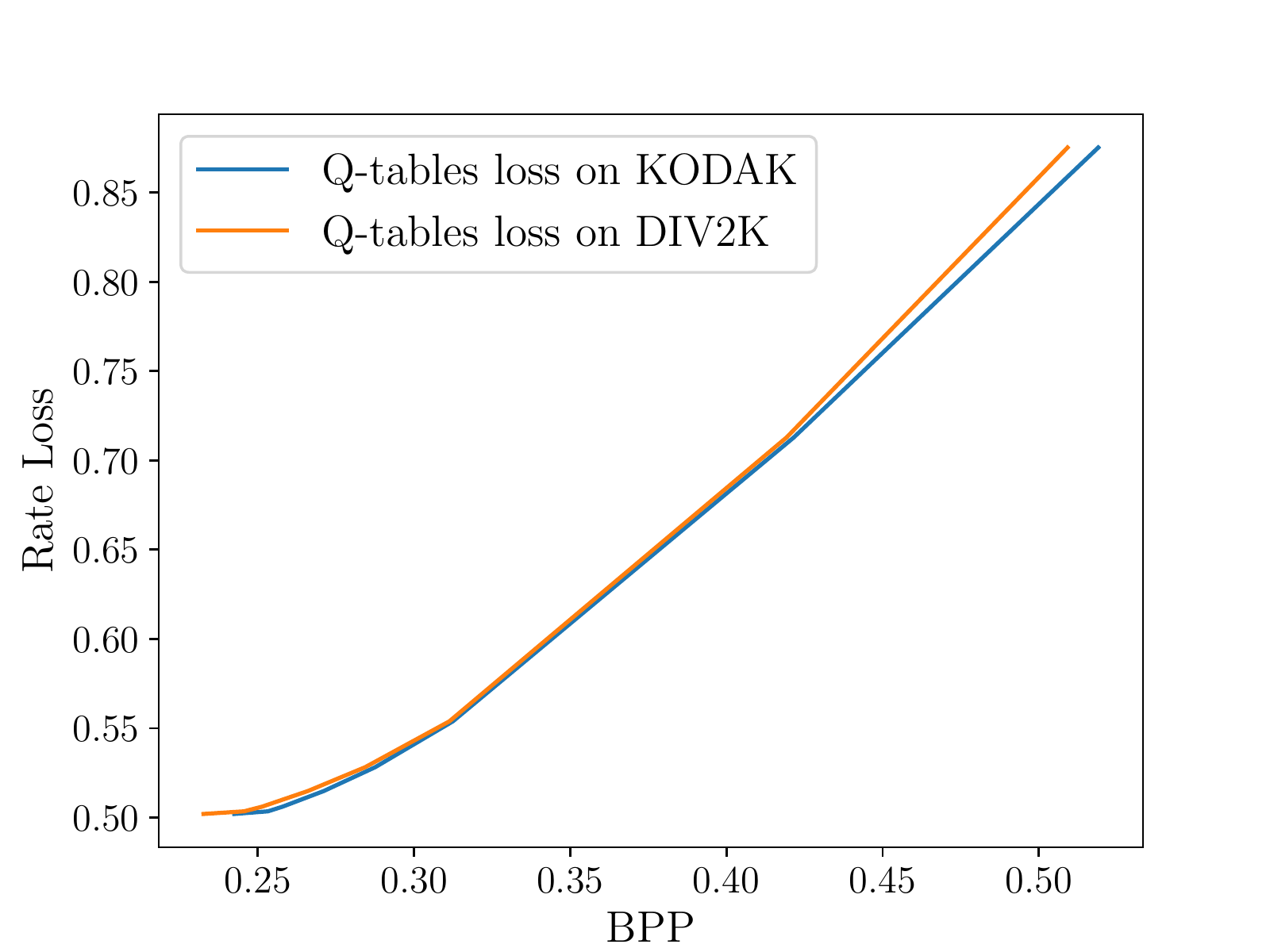}
\caption{Loss on quantization tables}
\end{subfigure}
\hfil
\begin{subfigure}[b]{.32\linewidth}
\includegraphics[width=\linewidth, trim={5 0 40 20}, clip=true]{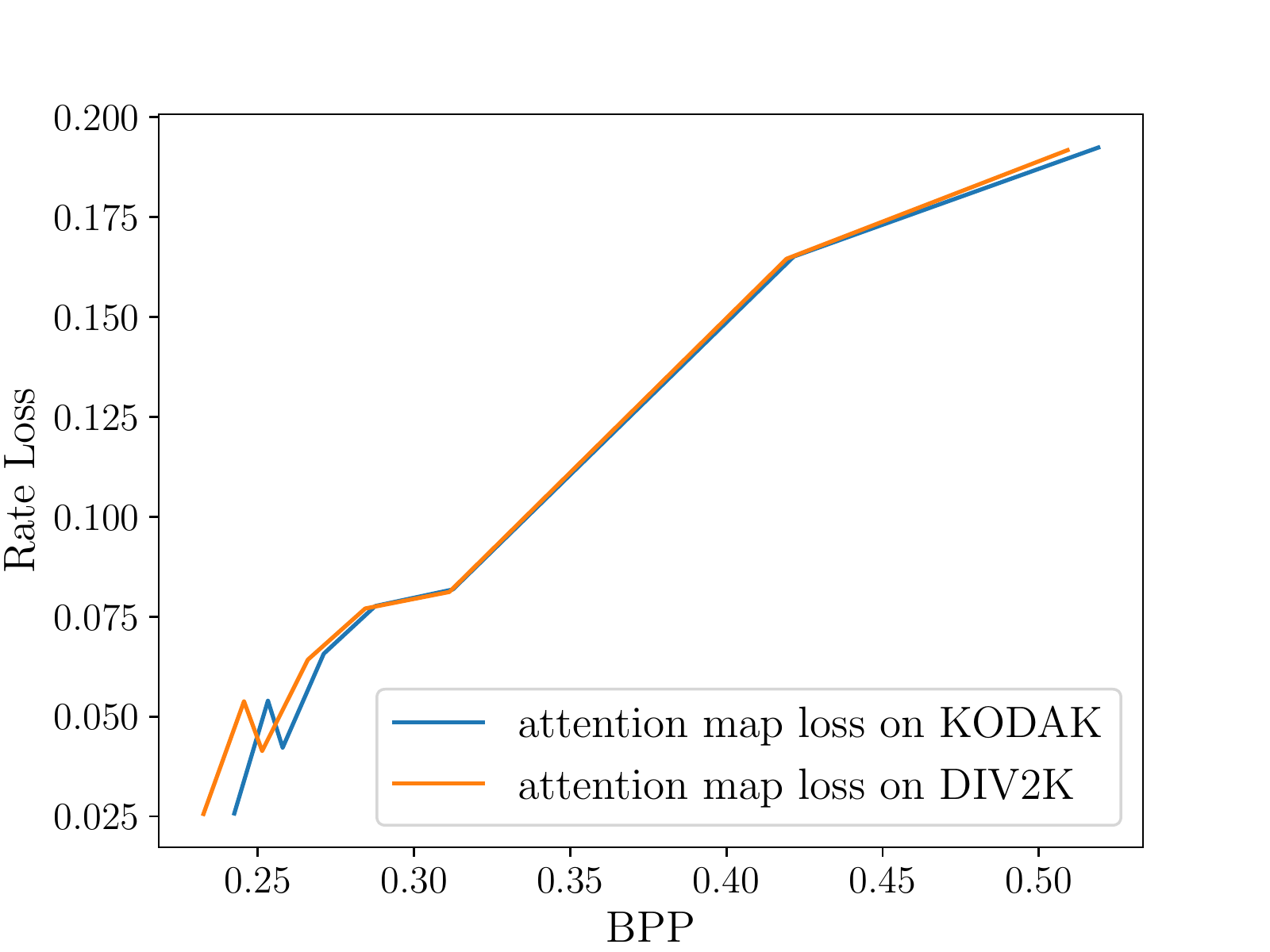}
\caption{Loss on attention maps}
\end{subfigure}
\hfil
\begin{subfigure}[b]{.32\linewidth}
\includegraphics[width=\linewidth, trim={5 0 40 20}, clip=true]{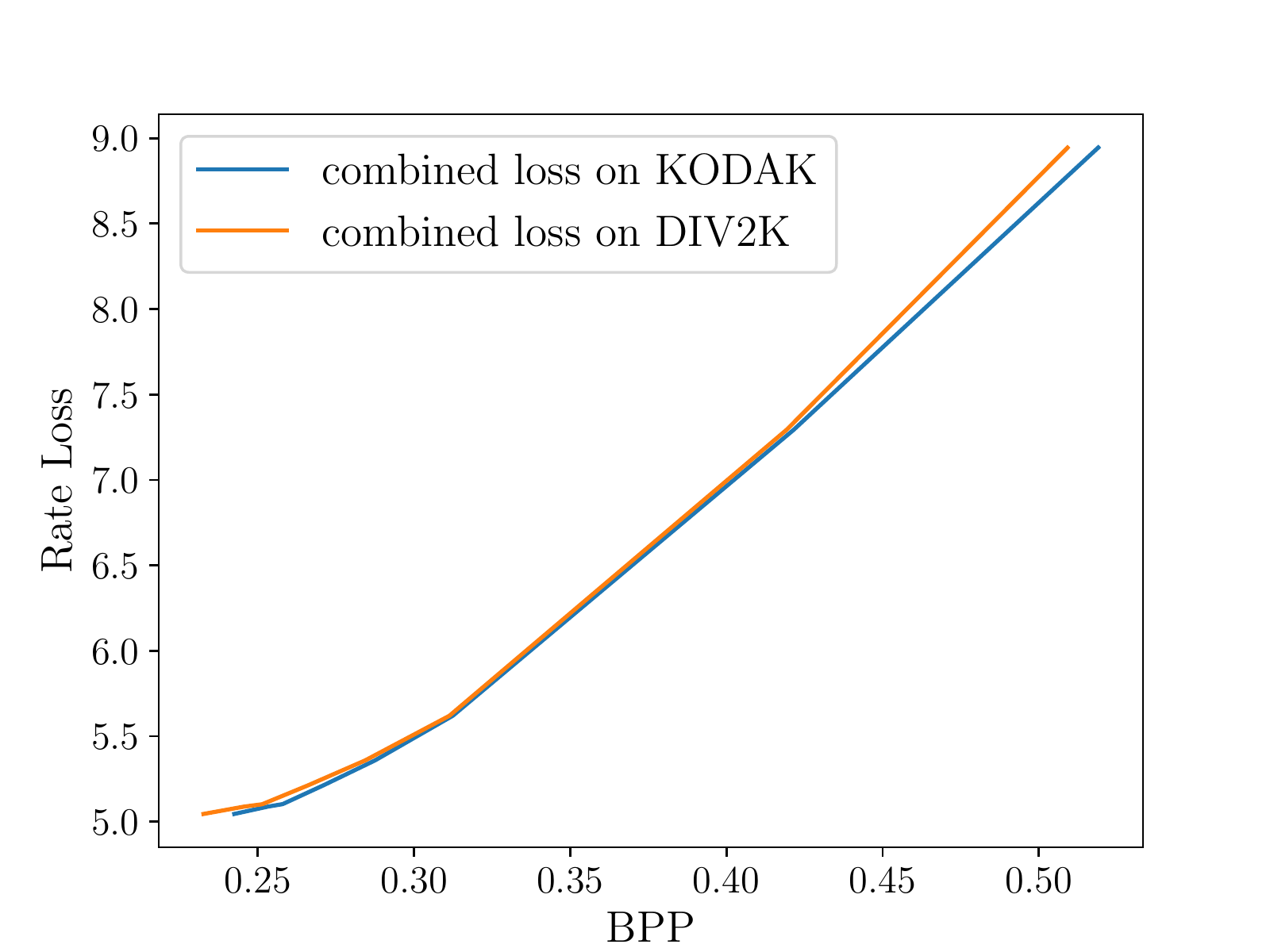}
\caption{Combined rate loss}
\end{subfigure}

\caption{Comparing our proposed rate loss to the true bpp. The loss terms are combined with the weights $\alpha=10$ (Q-tables), $\beta=1$ (attention maps).}
\label{fig:rate_loss}
\end{figure}

Since we cannot use the true bits per pixel of an image in the optimization we propose an alternative rate loss in \myeqref{rate_loss}. For the evaluation we use a model trained with the hyperparameters $\alpha=10$, $\beta=1$, $\gamma=500$. In \myfigref{fig:rate_loss}, we show the rate loss components and the combination of both for the DIV2K and Kodak dataset. Except for one outlier in the attention map loss, both components show a monotonically increasing curve, hence they are suitable to be used as a proxy to the true bit rate. Combining both loss terms with the weights $\alpha=10$, $\beta=1$ it is evident that the final rate loss used for optimization is dominated by the quantization table loss. We choose so because the majority of the entropy reduction should be achieved through the quantization tables.

\begin{figure}[!t]
\captionsetup[subfigure]{font=scriptsize}
\centering
\begin{subfigure}[b]{.32\linewidth}
\includegraphics[width=\linewidth]{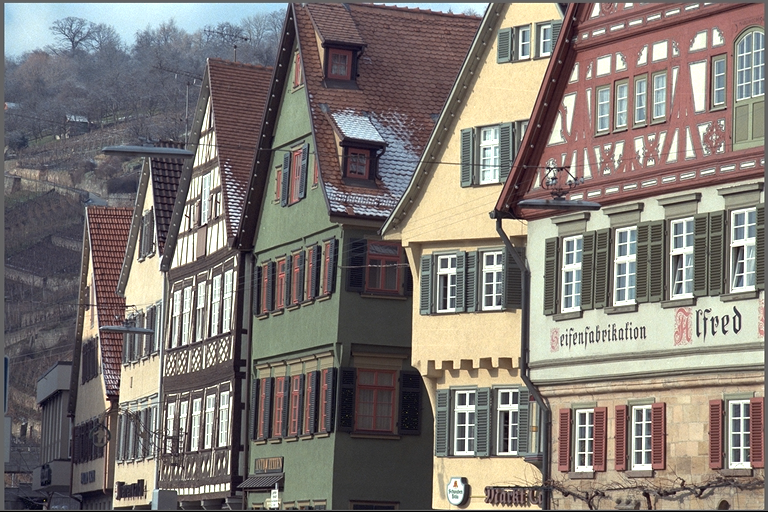}
\caption{original}
\end{subfigure}
\hfil
\begin{subfigure}[b]{.32\linewidth}
\includegraphics[width=\linewidth]{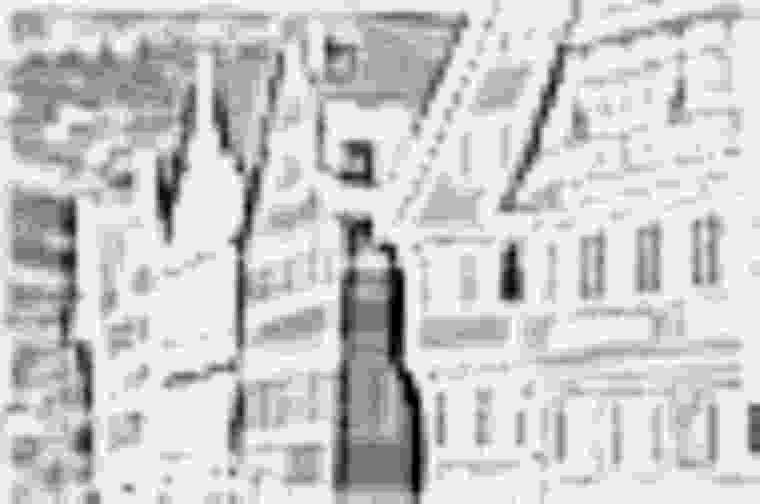}
\caption{luminance attention map}
\label{luma}
\end{subfigure}
\hfil
\begin{subfigure}[b]{.32\linewidth}
\includegraphics[width=\linewidth]{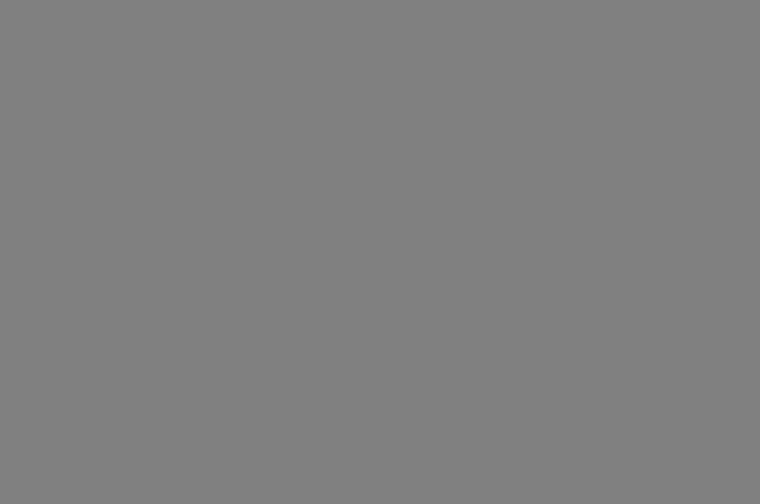}
\caption{chrominance attention map}
\label{chroma}
\end{subfigure}
\begin{subfigure}[b]{.32\linewidth}
\includegraphics[width=\linewidth]{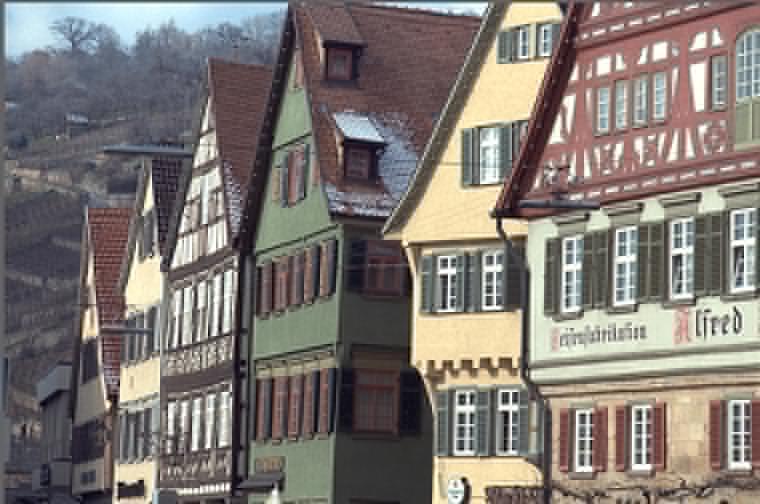}
\caption{smoothed}
\label{fig:smoothed}
\end{subfigure}
\hfil
\begin{subfigure}[b]{.32\linewidth}
\includegraphics[width=\linewidth]{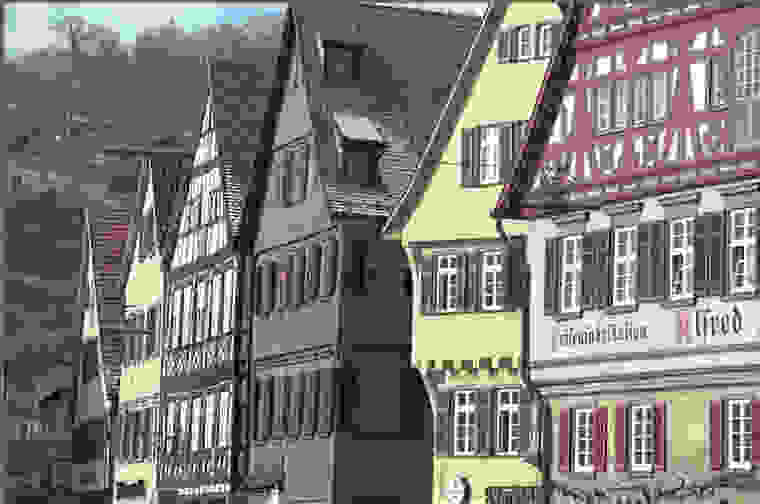}
\caption{JPEG ($0.337$ bpp)}
\end{subfigure}
\hfil
\begin{subfigure}[b]{.32\linewidth}
\includegraphics[width=\linewidth]{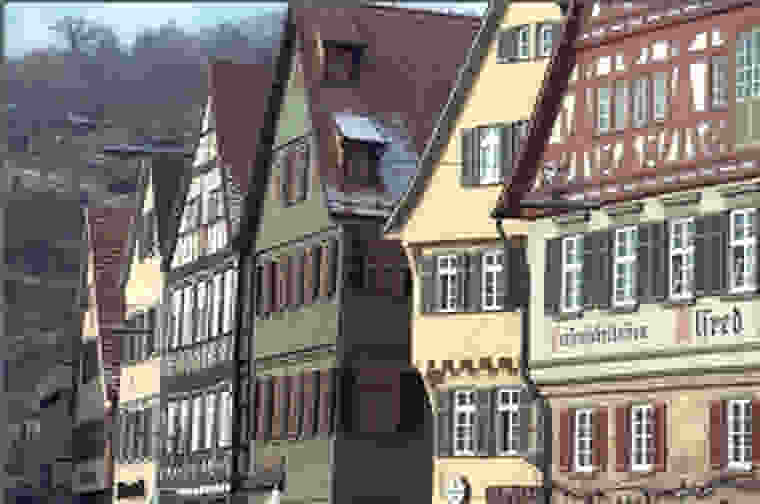}
\caption{ours ($0.321$ bpp)}
\end{subfigure}
\caption{Visualizing attention maps and smoothing output.}
\label{fig:attention_map}
\end{figure}

\subsection{Impact of the Attention Network}
In this section we want to visually show the effect of attention based editing to interpret its benefits. We use the network with attention trained with $\alpha=10, \beta=1, \gamma = 500$ and evaluate without rounding in the quantization step to only show the modifications through the attention maps. For visualization of the attention maps we average over all $8 \times 8$ DCT frequencies and show the spatial dimensions.  For better visibility we also normalize the attention maps.

Looking at \myfigref{fig:smoothed} shows the smoothed output of the attention based editing. We can generally see that fine details are reduced while strong edges are retained. \myfigref{luma} shows the attention map for the luminance channel and we can see that prominent edges (\eg \ roof line) and textures (\eg \  text on the building) get a higher attention score. Areas in the background and planes with few contrast edges get assigned a lower importance and are smoothed more. In the chrominance attention map in \myfigref{chroma} we do not see any spatial adaptation. Comparing the final compression result of our network to the JPEG baseline shows that there is slightly less detail in the letter on the building, however overall the image looks more pleasing with significantly better color retention. The benefit of the smoothing is that the quantization table can contain smaller values while still achieving similar bit rates. Smaller quantization table values translate to more levels in the image that overall leads to better colors.

\begin{figure}[!t]
\centering
\captionsetup[subfigure]{labelformat=empty}
\begin{subfigure}[b]{.32\linewidth}
\includegraphics[width=\linewidth, trim={5 0 40 10}, clip=true]{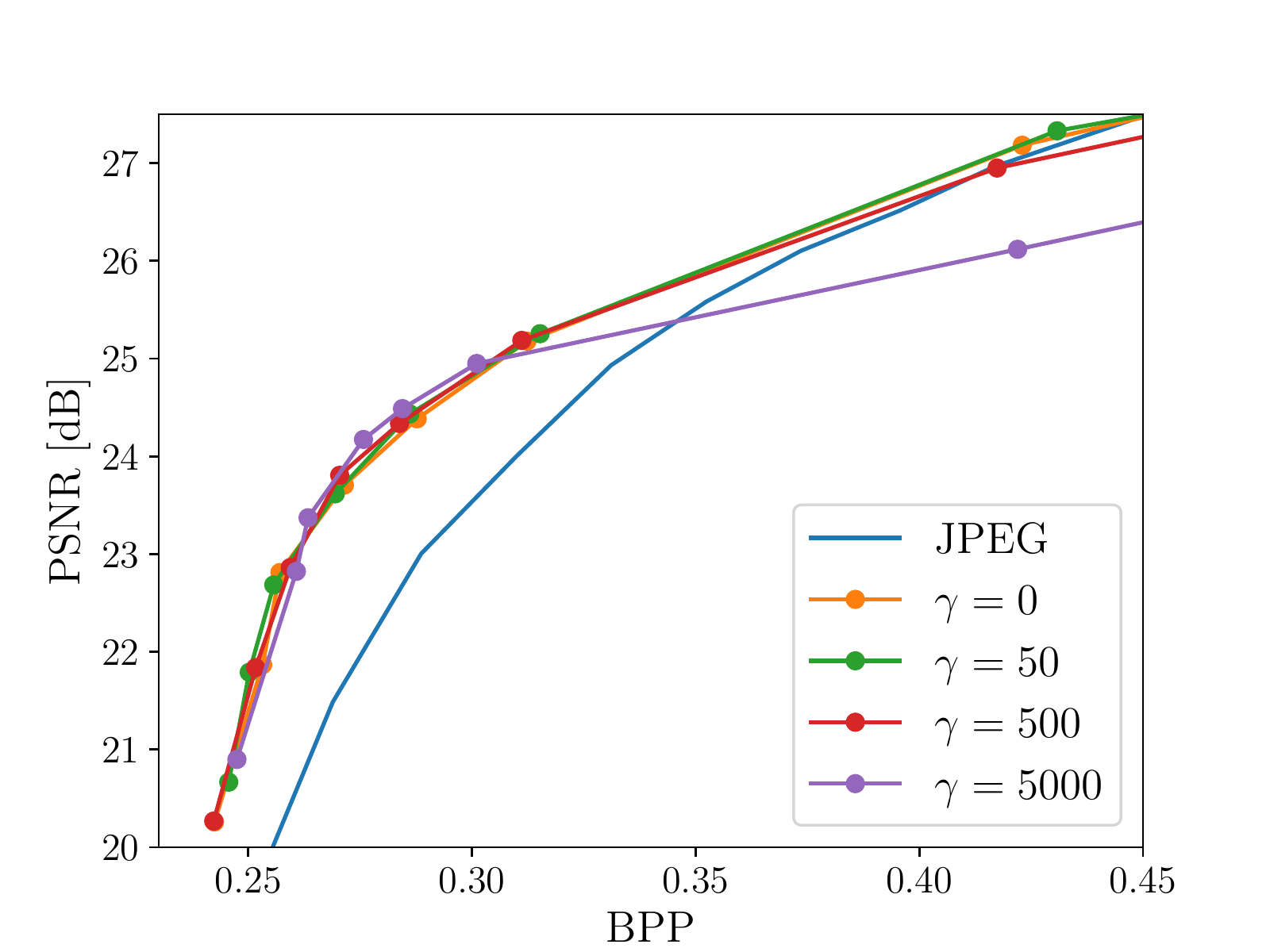}
\caption{PSNR on Kodak}
\end{subfigure}
\hfil
\begin{subfigure}[b]{.32\linewidth}
\includegraphics[width=\linewidth, trim={5 0 40 20}, clip=true]{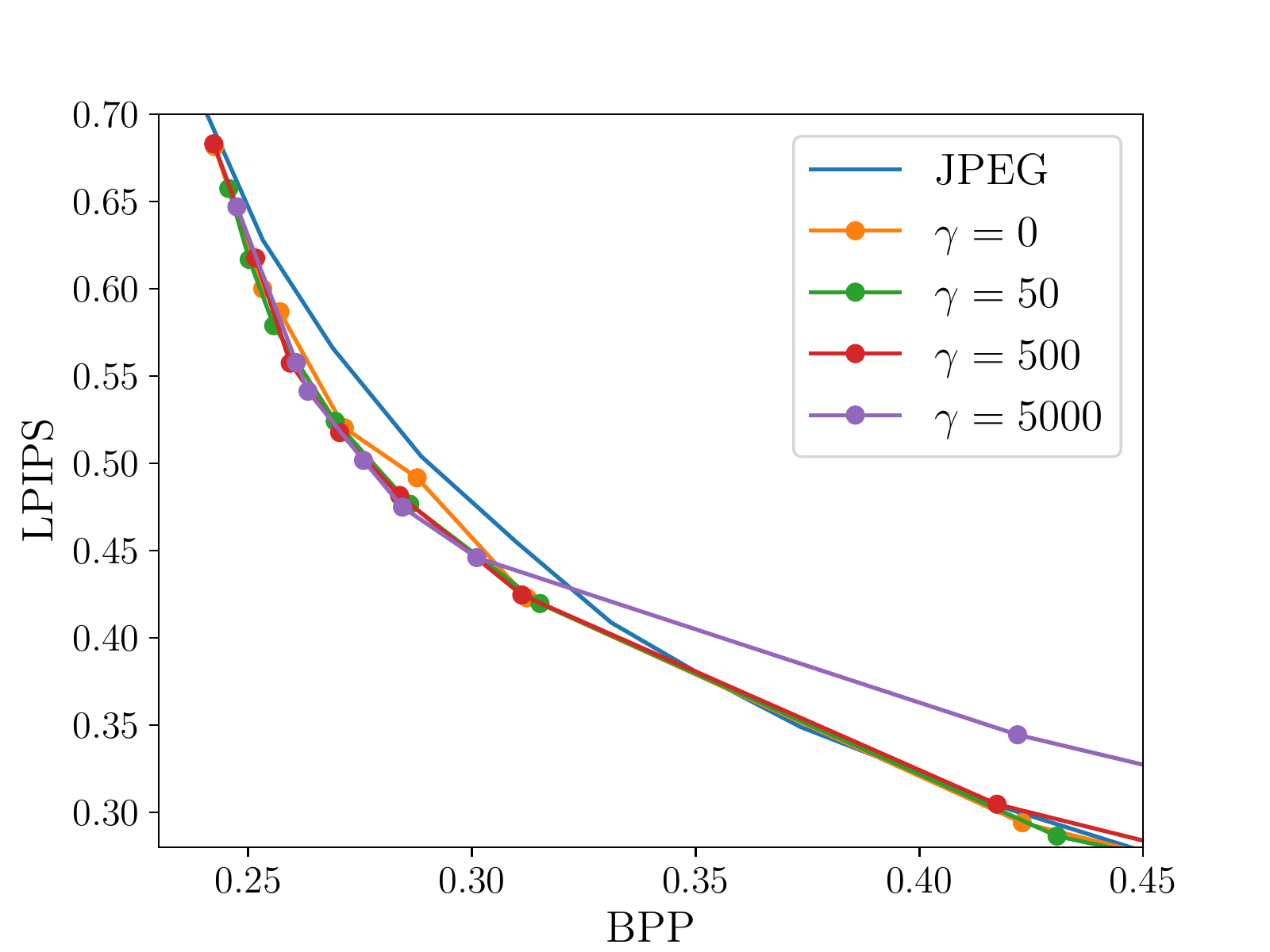}
\caption{LPIPS on Kodak}
\end{subfigure}
\hfil
\begin{subfigure}[b]{.32\linewidth}
\includegraphics[width=\linewidth, trim={5 0 40 20}, clip=true]{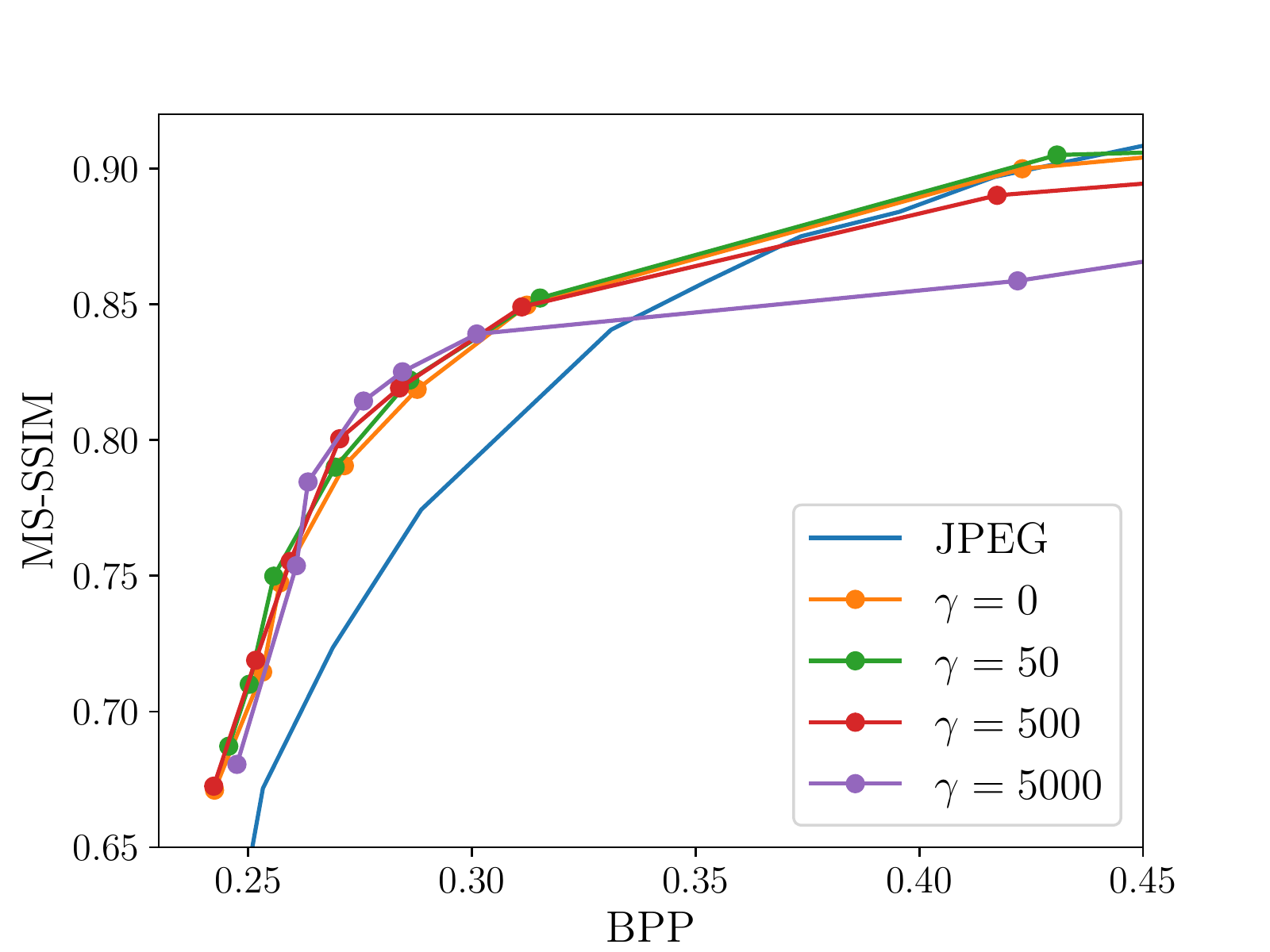}
\caption{MS-SSIM on Kodak}
\end{subfigure}
\begin{subfigure}[b]{.32\linewidth}
\includegraphics[width=\linewidth, trim={5 0 40 20}, clip=true]{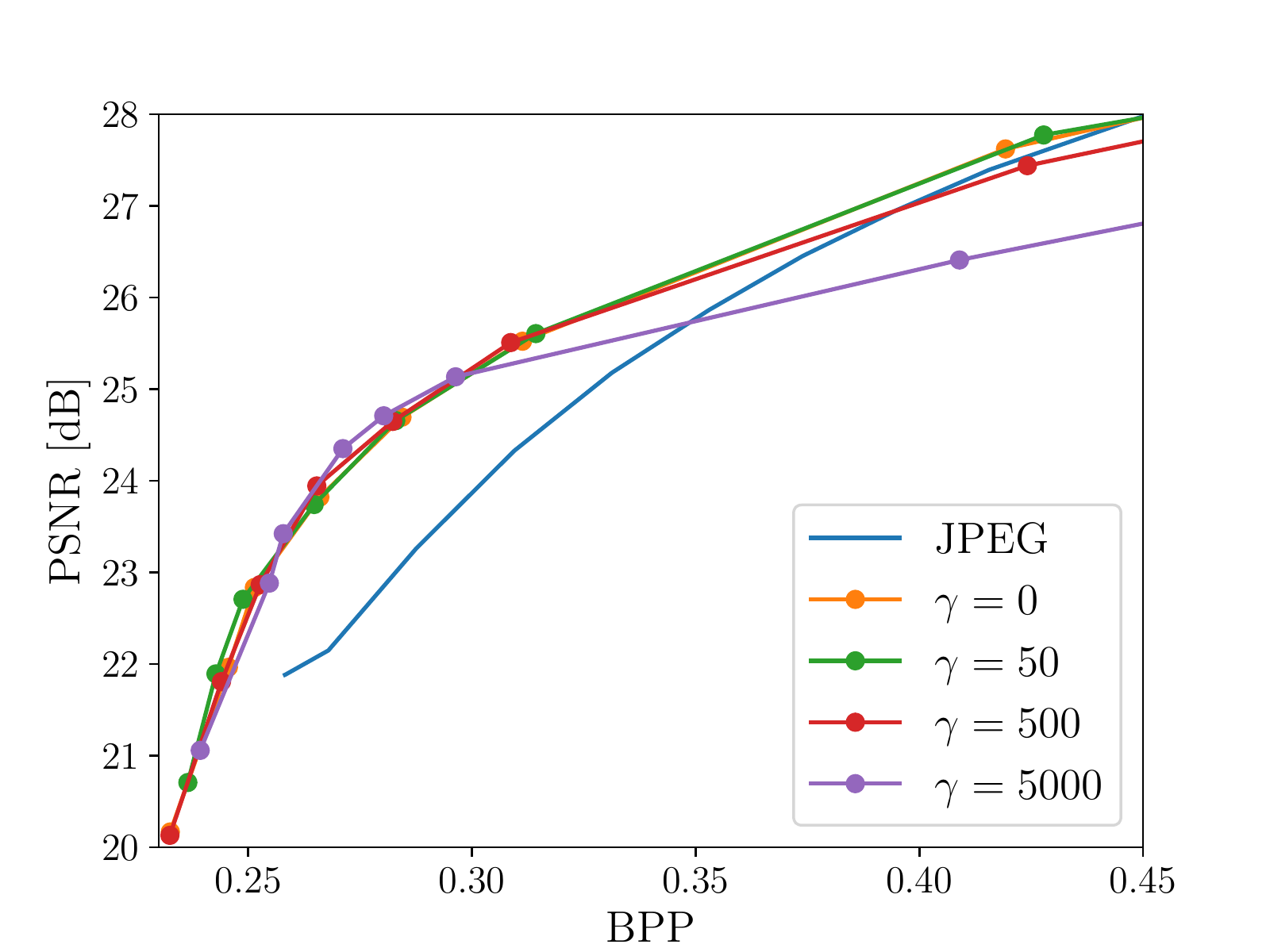}
\caption{PSNR on DIV2K}
\end{subfigure}
\hfil
\begin{subfigure}[b]{.32\linewidth}
\includegraphics[width=\linewidth, trim={5 0 40 20}, clip=true]{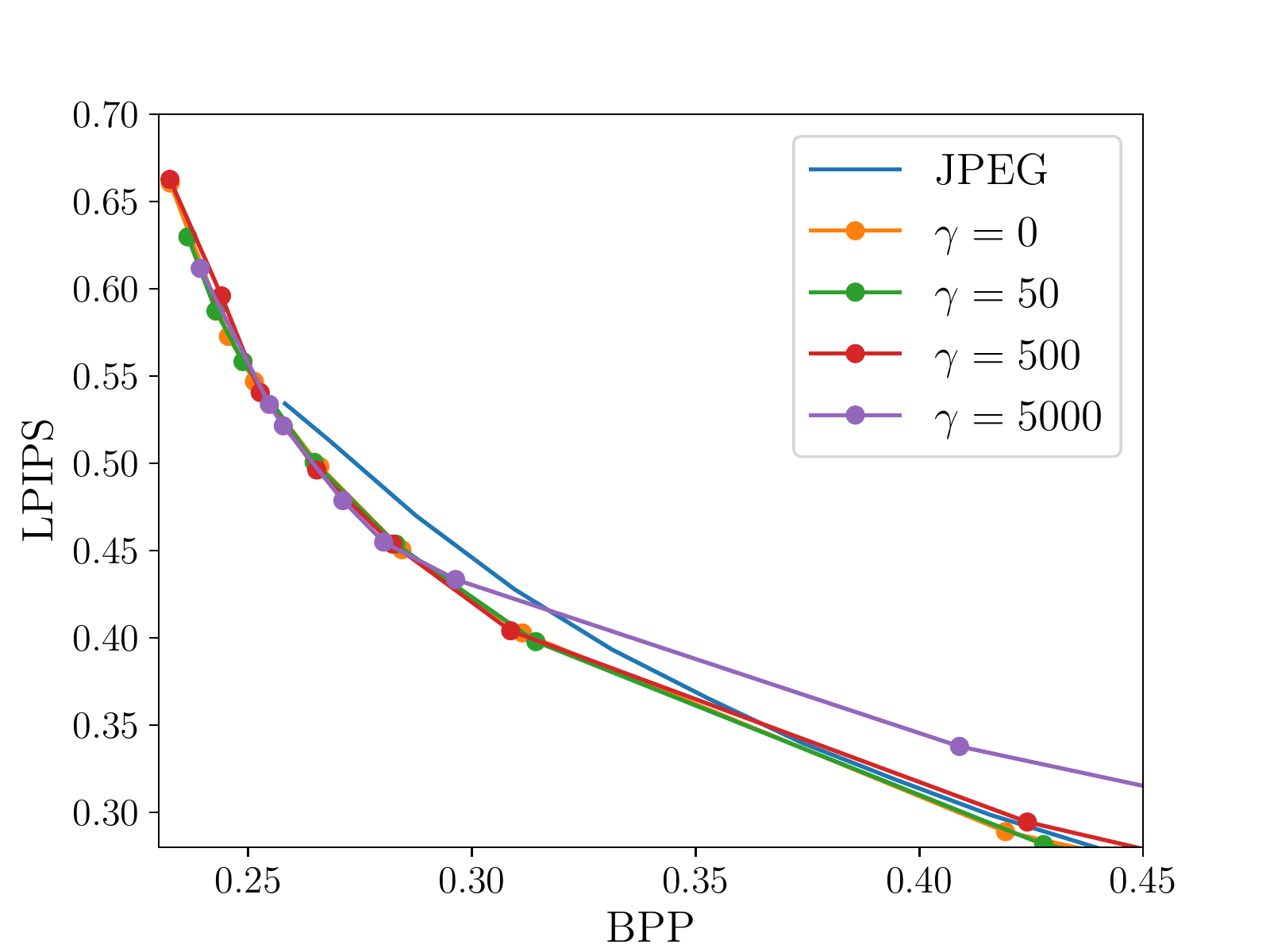}
\caption{LPIPS on DIV2K}
\end{subfigure}
\hfil
\begin{subfigure}[b]{.32\linewidth}
\includegraphics[width=\linewidth, trim={5 0 40 20}, clip=true]{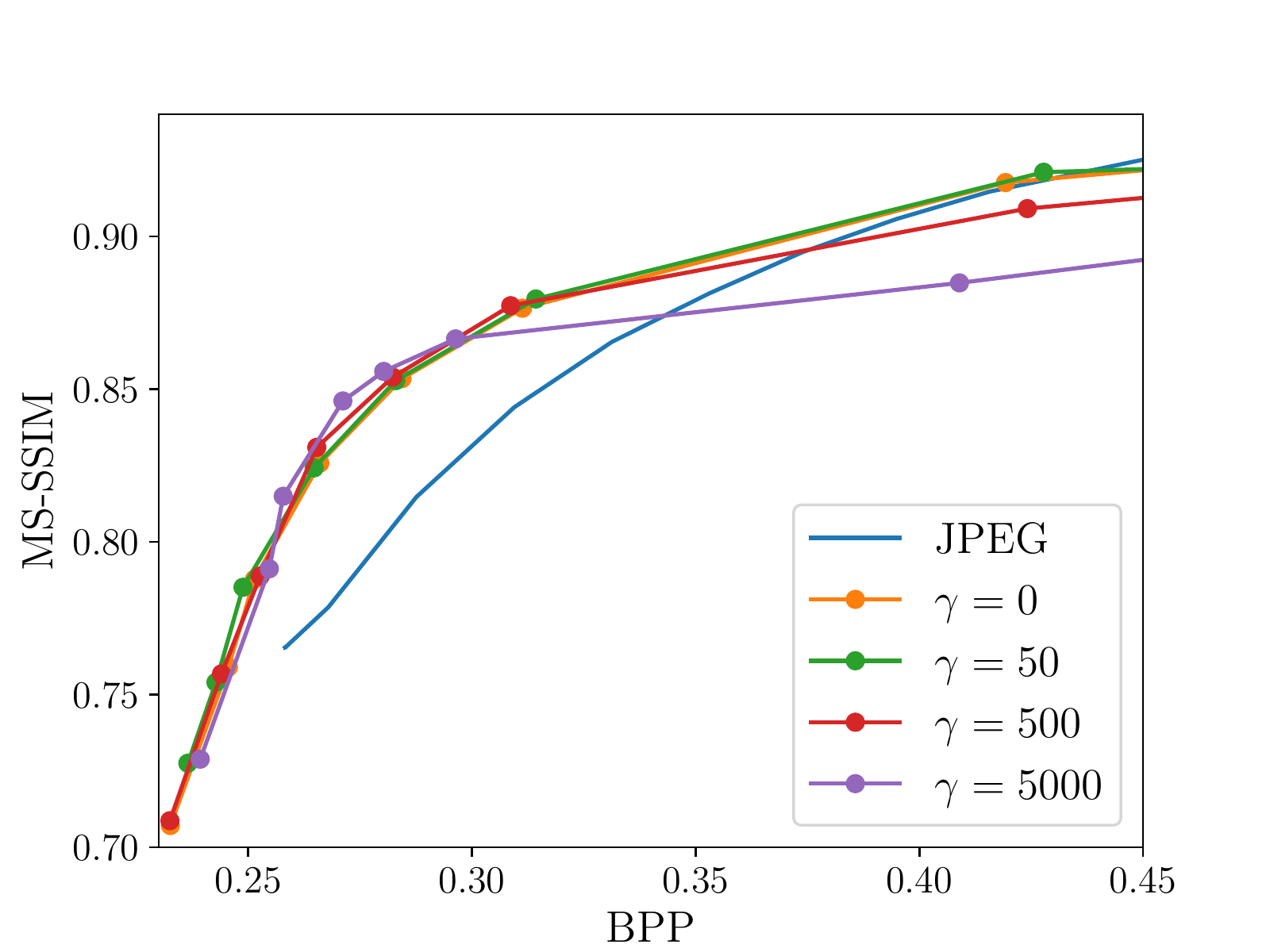}
\caption{MS-SSIM on DIV2K}
\end{subfigure}
\caption{Ablation study for varying the LPIPS loss weight $\gamma \in \{0, 50, 500, 5000\}.$}
\label{fig:lpips}
\end{figure}

\subsection{Impact of the Perceptual Loss}
In this experiment we show how the addition of a perceptual loss term effects the compression performance. Setting the hyperparameter of the perceptual loss to $\gamma=500$ leads to an approximately even loss contribution of MSE and LPIPS. Setting $\gamma=50, 5000$ we put less or more emphasis on the LPIPS loss term. The other weight hyperparameters stay as before: $\alpha=10$, $\beta=1$.

\begin{figure}[!t]
\centering

\captionsetup[subfigure]{labelformat=empty, font=small}
\begin{subfigure}[b]{.15\linewidth}
\caption{original}
\includegraphics[width=\linewidth, trim={350 150 250 200}, clip=true ]{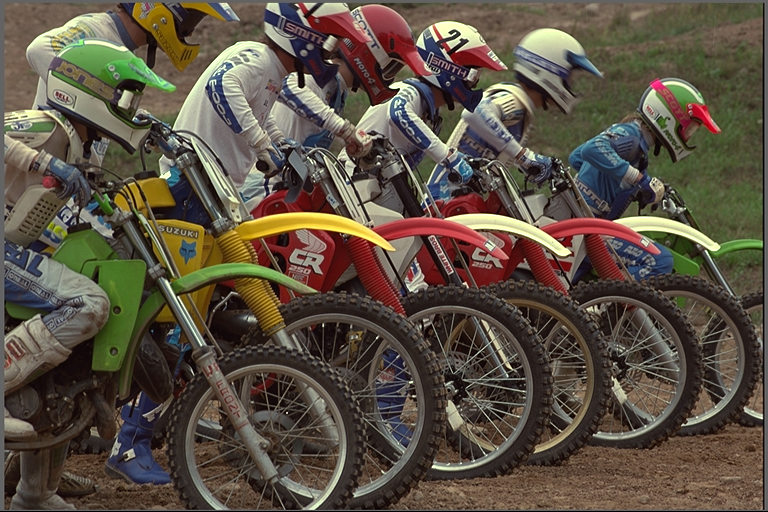}
\caption{$16.4$ bpp}
\end{subfigure}
\hfil
\begin{subfigure}[b]{.15\linewidth}
\caption{JPEG}
\includegraphics[width=\linewidth, trim={350 150 250 200}, clip=true ]{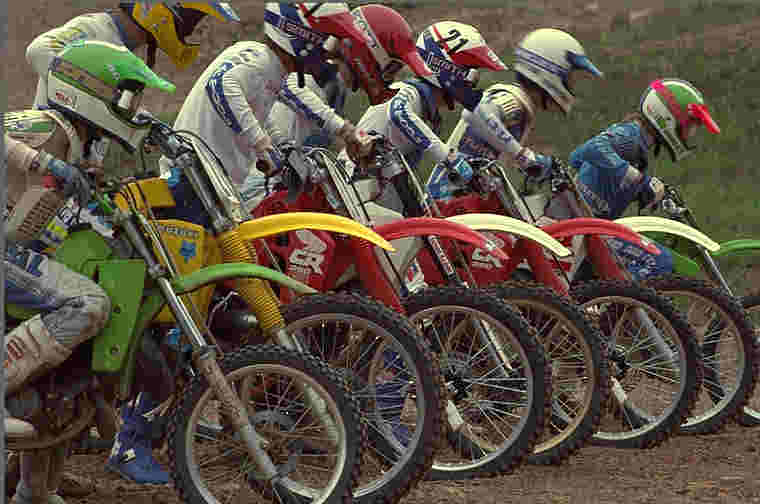}
\caption{$0.636$ bpp}
\end{subfigure}
\hfil
\begin{subfigure}[b]{.15\linewidth}
\caption{$\gamma = 0$}
\includegraphics[width=\linewidth, trim={350 150 250 200}, clip=true ]{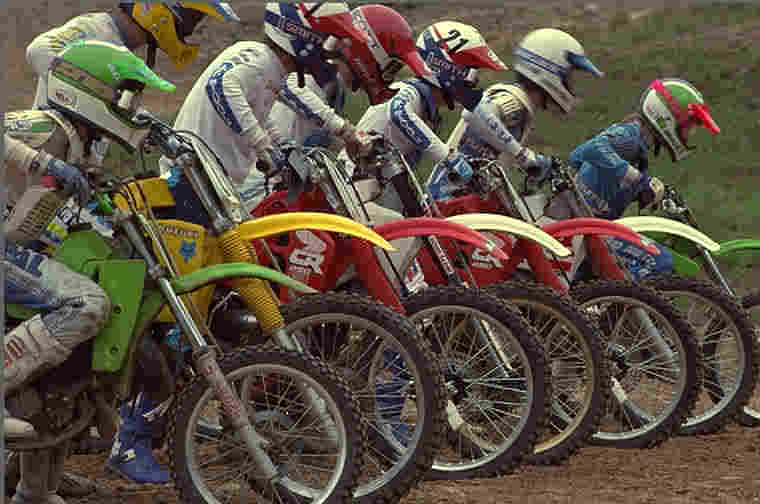}
\caption{$0.616$ bpp}
\end{subfigure}
\hfil
\begin{subfigure}[b]{.15\linewidth}
\caption{$\gamma = 50$}
\includegraphics[width=\linewidth, trim={350 150 250 200}, clip=true ]{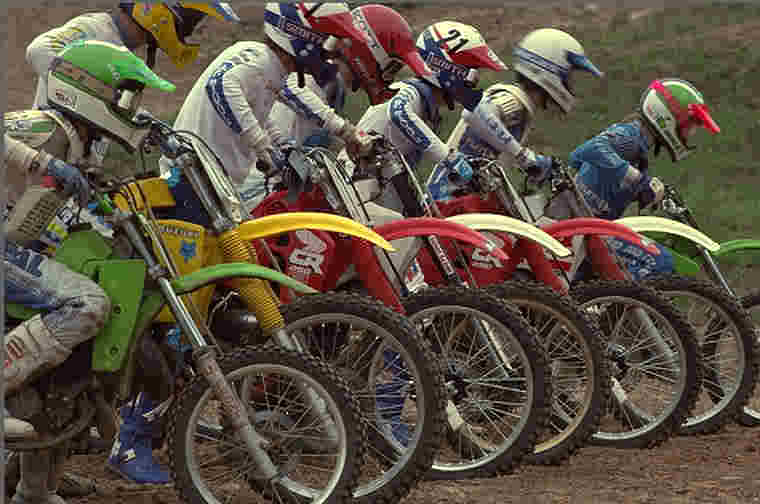}
\caption{$0.630$ bpp}
\end{subfigure}
\hfil
\begin{subfigure}[b]{.15\linewidth}
\caption{$\gamma = 500$}
\includegraphics[width=\linewidth, trim={350 150 250 200}, clip=true ]{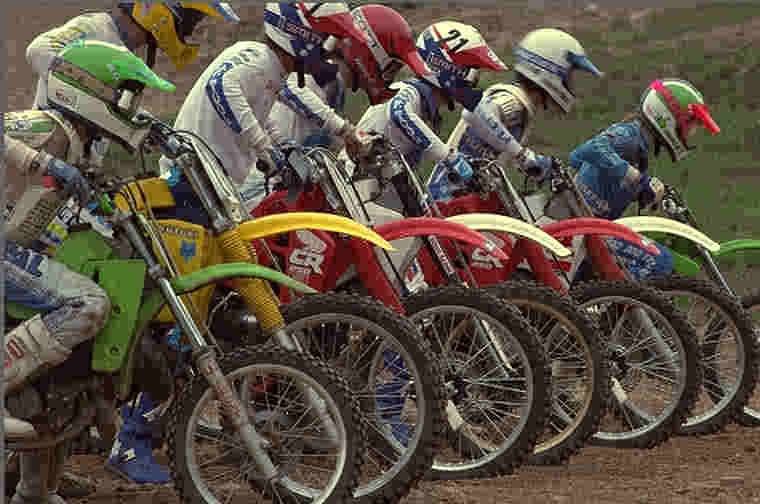}
\caption{$0.603$ bpp}
\end{subfigure}
\hfil
\begin{subfigure}[b]{.15\linewidth}
\caption{$\gamma = 5000$}
\includegraphics[width=\linewidth, trim={350 150 250 200}, clip=true ]{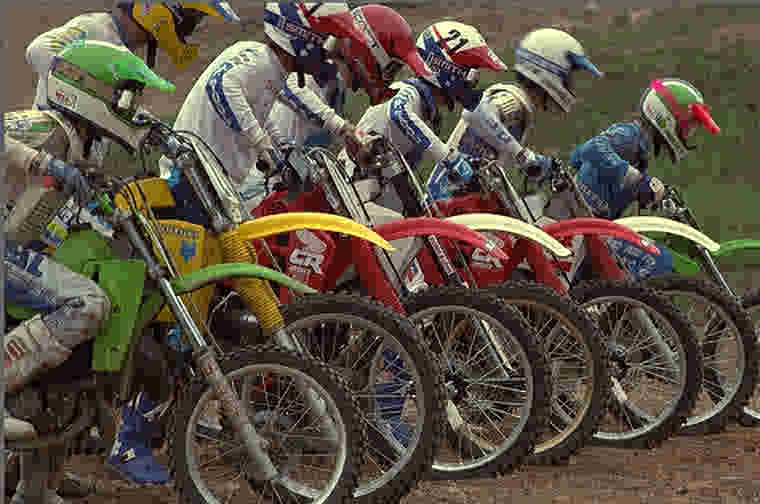}
\caption{$0.555$ bpp}
\end{subfigure}
\caption{Visual quality at different LPIPS weights (cropped view).}
\label{fig:lpips_im_crop}
\end{figure}

When looking at \myfigref{fig:lpips} we do not see a big difference in performance between setting the LPIPS weight to $\gamma = 0$, $\gamma = 50$ or $\gamma = 500$. The general trend we see is that higher $\gamma$ show a sweet spot
between $0.25$-$0.3$ bpp, but have slightly worse performance on the low and high end of the reported range. This is especially visible for $\gamma = 5000$. When comparing the image output in \myfigref{fig:lpips_im_crop}, we see the best detail retention and color reproduction in the wheel for $\gamma = 50$ and $\gamma = 500$. Both have clearly less color artifacts than the JPEG baseline. This is especially visible around the tires where the JPEG baseline produces purple color artifacts. For $\gamma = 5000$ we see a slightly worse performance than for a lower settings of $\gamma$. Although the image looks slightly sharper, the blockiness is also more apparent.

\subsection{Refinement on Kodak}
In a further experiment we started off with the network with attention that was already trained on the Hasinoff \cite{hasinoff} dataset with the hyperparameters set to $\alpha=10$, $\beta=1$, $\gamma=500$. We use images 13-24 from the Kodak \cite{kodak} dataset to train the network for another 10000 steps, resulting in 30000 steps in total. This is done for each setting of the tradeoff parameter $\lambda$. The testing is performed on images 1-12 in the dataset.

The goal of refining the network on part of the test set, is to see how much the compression performance benefits when the distribution of train and test images is closer. In other words, large improvements suggest that the network does not generalize well from the Hasinoff \cite{hasinoff} to the Kodak \cite{kodak} dataset.
Comparing the metrics shown in \myfigref{fig:refine} we see almost no improvement at low bit rates. At higher bit rates, the network benefits increasingly from the refinement. This is especially visible in the MS-SSIM metric. After the refinement, the network achieves similarity metrics that are roughly equal to the JPEG baseline above $0.4$ bpp, whereas previously the performance was inferior for these higher bitrates. 

\subsection{Single Image Optimization}
In this experiment we explore the limit of the achievable compression of our network with attention  by optimizing it for each individual test image. We do so by following the same training procedure as usual but since we only have a single image in the training set, we use a batch size of 1 and train directly on the full image (without random cropping as before).We keep all hyperparameters the same as for the refinement. In total we train our model for 6 values of the tradeoff parameter $\lambda \in [10
^{-3}, 10^{-1} ]$ for each of the 24 images in the Kodak dataset. This results in a total of $144$ models that we then evaluate for the particular image that each model was trained on. As before we average the measured metrics over each setting of $\lambda$ to get an overall rate-distorion curve.

\begin{figure}[!t]
\centering
\captionsetup[subfigure]{labelformat=empty}
\begin{subfigure}[b]{.32\linewidth}
\includegraphics[width=\linewidth, trim={5 0 40 20}, clip=true]{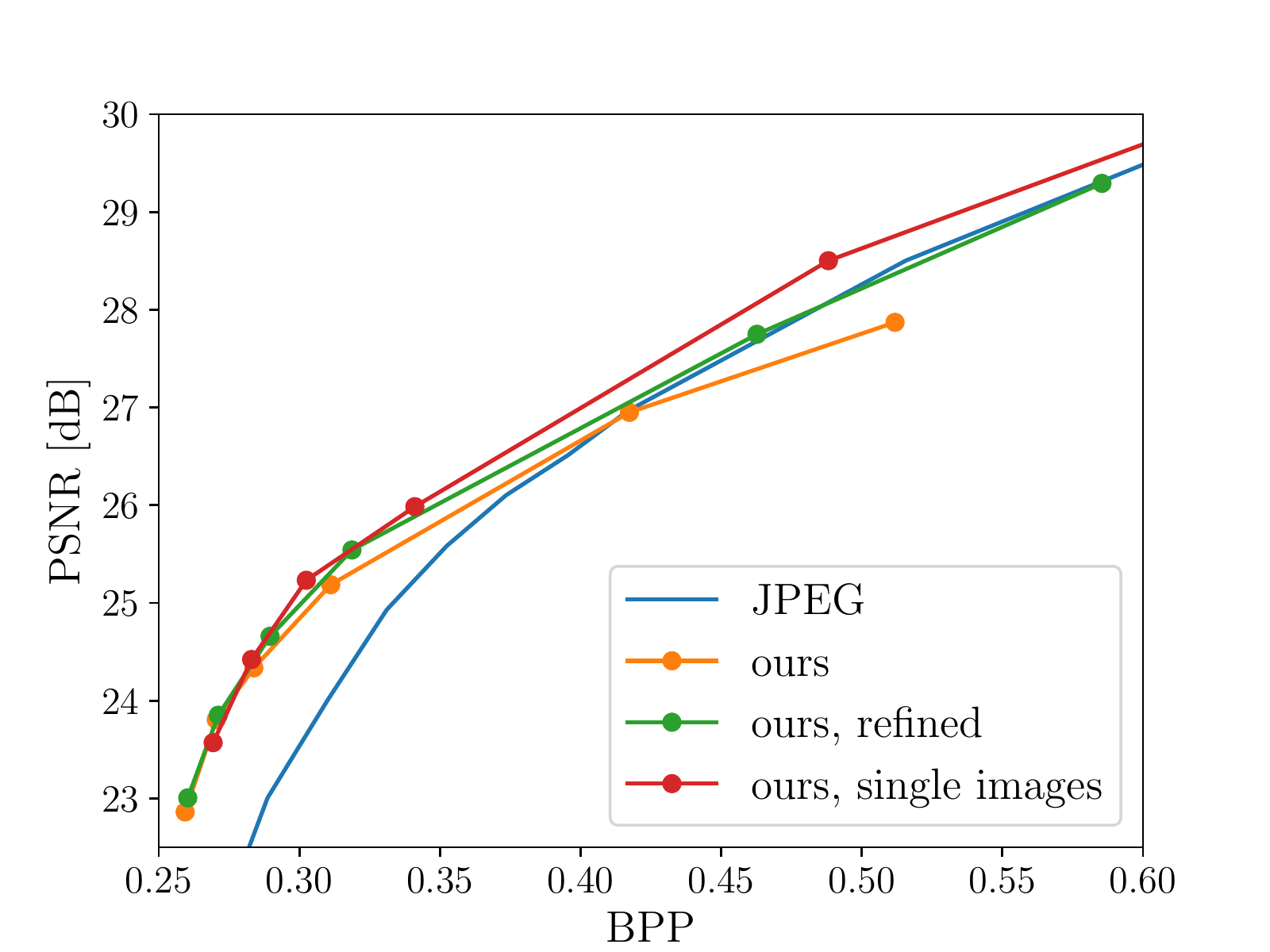}
\caption{PSNR on Kodak}
\end{subfigure}
\hfil
\begin{subfigure}[b]{.32\linewidth}
\includegraphics[width=\linewidth, trim={5 0 40 20}, clip=true]{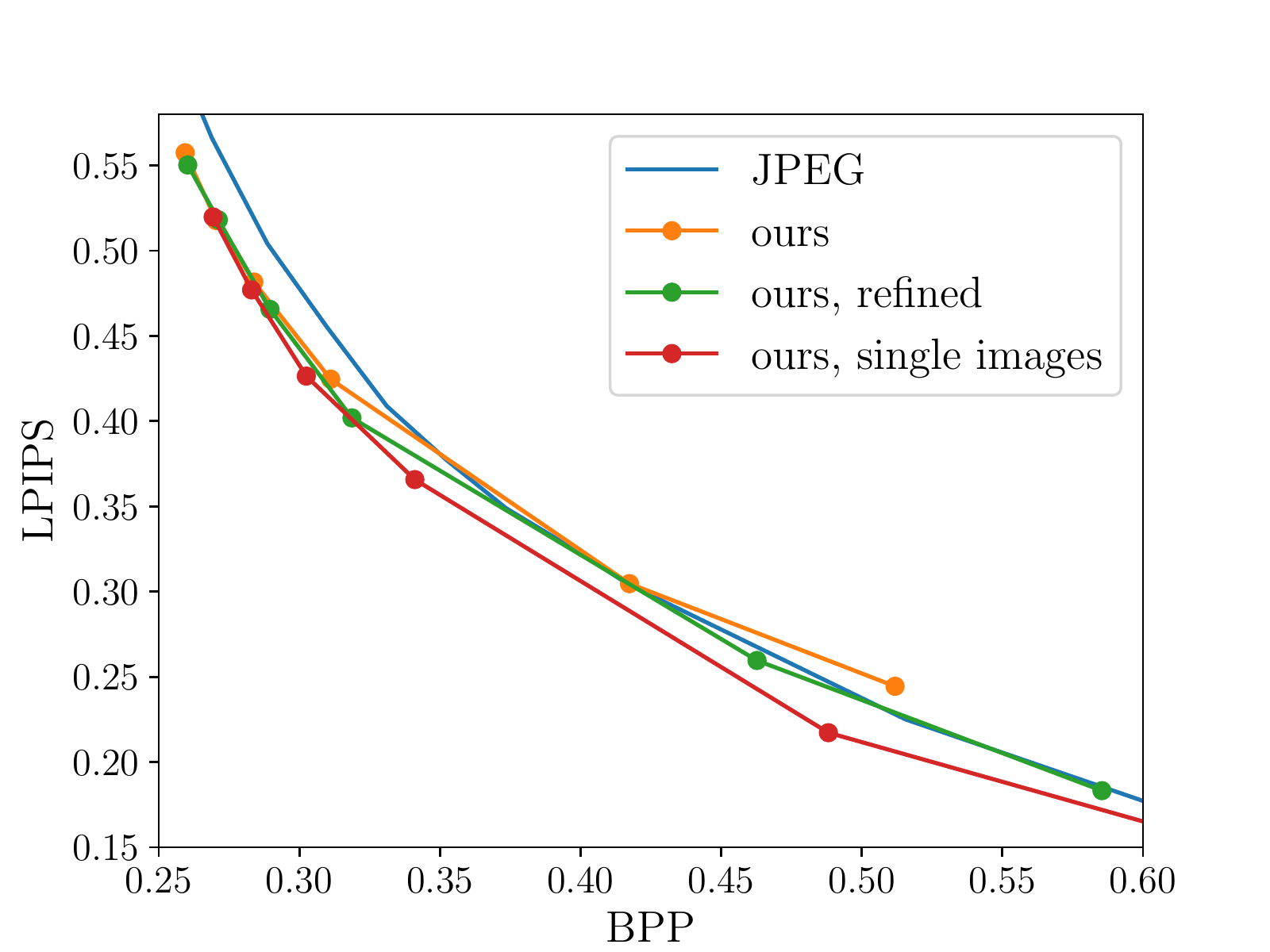}
\caption{LPIPS on Kodak}
\end{subfigure}
\hfil
\begin{subfigure}[b]{.32\linewidth}
\includegraphics[width=\linewidth, trim={0 0 40 20}, clip=true]{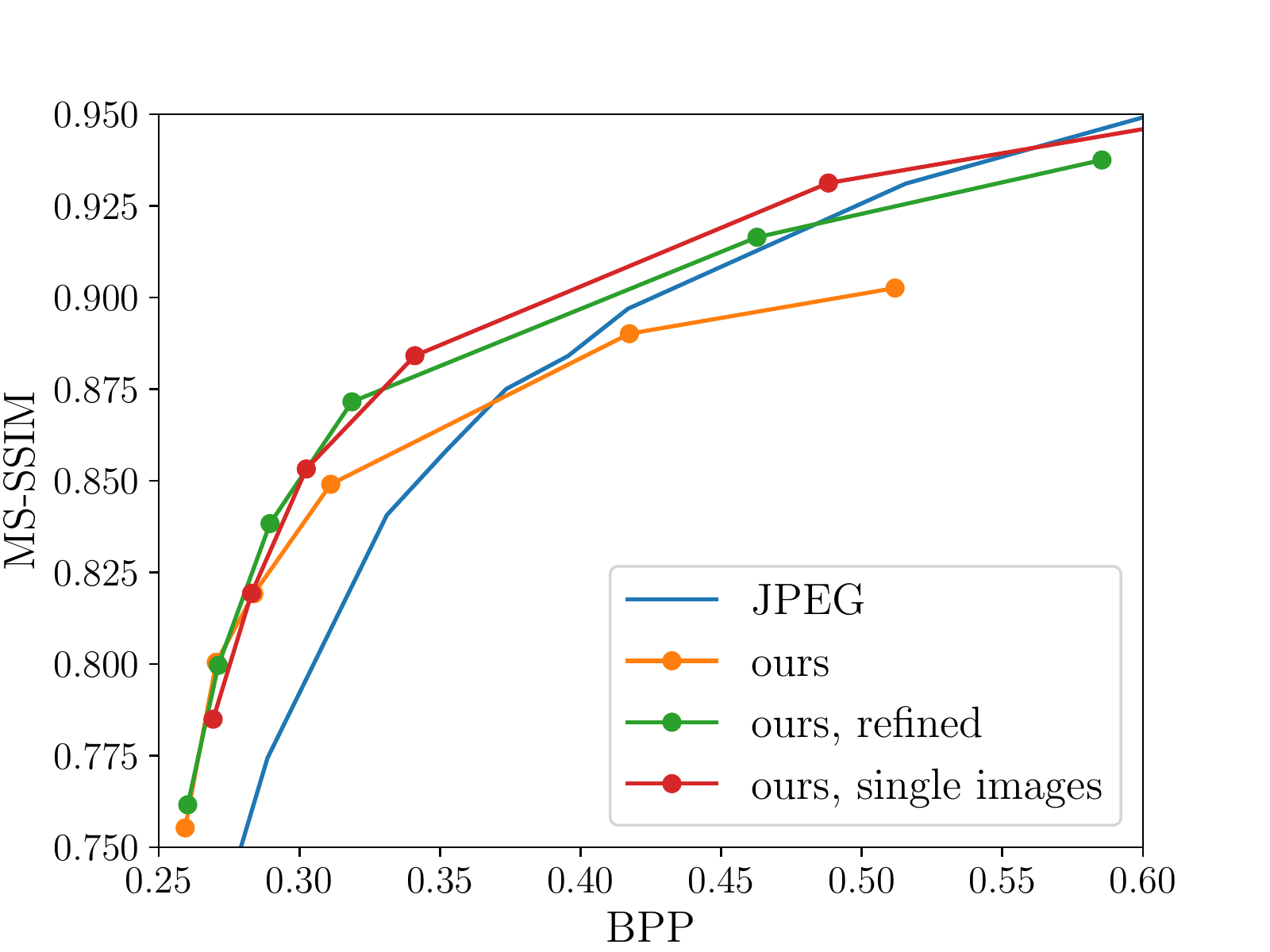}
\caption{MS-SSIM on Kodak}
\end{subfigure}

\caption{Comparing the rate-distortion performance of our solution when refined on part of the Kodak test set and when trained for single images.}
\label{fig:refine}
\end{figure}

Looking at the result shown in \myfigref{fig:refine} shows that the networks trained on single images outperform the refined and unrefined network at bit rates larger than $0.3$ bpp as to be expected. For very low bit rates all three approaches perform roughly the same. These findings support the belief that the worse performance of the normal model at higher bit rates, stems from the lack of generalization. In fact, training and testing on single images does not require any generalization at all, explaining the higher scores.
\section{Conclusion}
We have shown in our experiments that our approach to improving JPEG encoding through attention based smoothing and learned quantization tables leads to better compression for low bit rates in terms of absolute deviation (MSE, PSNR) and perceptual metrics (LPIPS, MS-SSIM). Most importantly, the improvemed encoder remains entirely compatible with any standard JPEG decoder.  Working at low bit rates, our solution still produces significant blocking artifacts that are simply inherent to the JPEG algorithm and cannot be avoided. We were however able to retain significantly better colors than the standard JPEG implemention. We have also shown that we can achieve this improvements without directly estimating the entropy of the DCT coefficients, but by regularizing the attention maps and the learned quantization tables. Our experiments on optimizing for single images have shown that individually learned quantization tables perform better at bit rates higher than $0.35$.
As a possible extension of our approach we suggest to additionally learn a network predicting the quantization tables for each image individually instead of learning global quantization tables that work for all images. This would make the quantization tables adaptive to each image, while the compression can still be done in a single forward evaluation.

Improving compression quality at low bit rates is relevant to all areas that require very small image files. A possible application are websites where the content should be loaded as fast as possible even when the internet connection is slow. The proposed solution can also be used to render previews that are updated with a higher quality version once they are fully loaded.

%
%
\bibliographystyle{splncs04}
\bibliography{bib.bib}
\end{document}